\documentclass{aa} % for a printer version
\newcommand{\gtsim}{\protect\raisebox{-0.5ex}{$\:\stackrel{\textstyle >}
        {\sim}\:$}}
\newcommand{\ltsim}{\protect\raisebox{-0.5ex}{$\:\stackrel{\textstyle <}
        {\sim}\:$}}

\usepackage{longtable}    
\usepackage{aalongtable}    
\usepackage{supertabular}    
\usepackage{graphicx}
\usepackage{epsfig}
\usepackage{rotating}
\usepackage{lscape}
\usepackage[authoryear]{natbib}
\bibpunct{(}{)}{;}{a}{}{,} % to follow the A&A style

\begin{document}

\title{Chemical pattern across the young associations ONC and OB1b\thanks{Based on {\sc flames/uves} 
observations collected at the Paranal Observatory (Chile). Program 082.D-0796(A).} }
   
\author{K. Biazzo \inst{1} \and S. Randich\inst{1} \and F. Palla\inst{1}}
\offprints{K. Biazzo}
\mail{kbiazzo@arcetri.astro.it}

\institute{INAF - Osservatorio Astrofisico di Arcetri, Largo E. Fermi 5, 50125 Firenze, Italy}

\date{Received / accepted }

\abstract
% Context:
{Abundances of iron-peak and $\alpha$-elements are poorly known in Orion, and the available measurements yield 
contradictory results.
}
% Aims:
{We measure accurate and homogeneous elemental abundances of the Orion subgroups ONC and OB1b, and search 
for abundance differences across the Orion complex.
}
% Methods:
{We present {{\sc flames/uves}} spectroscopic observations of 20 members of the ONC and OB1b. We measured radial 
velocity, veiling, effective temperature using two spectroscopic methods, and determined the chemical abundances 
of Fe, Na, Al, Si, Ca, Ti, and Ni using the code MOOG. We also performed a new consistent analysis of spectra 
previously analyzed by our group.
} 
% Results:
{We find three new binaries in the ONC, two in OB1b, and three non-members in OB1b (two of them most likely 
being OB1a/25~Ori members). Veiling only affects one target in the ONC, and the effective temperatures derived 
using two spectroscopic techniques agree within the errors. The ONC and OB1b are characterized by a small scatter 
in iron abundance, with mean [Fe/H] values of $-0.11\pm0.08$ and $-0.05\pm0.05$, respectively. We find a small 
scatter in all the other elemental abundances. We confirm that P1455 is a metal-rich star in the ONC. 
}
% Conclusions: 
{We conclude that the Orion metallicity is not above the solar value. The OB1b group might be slightly more 
metal-rich than the ONC; on the other hand, the two subgroups have similar almost solar abundances of iron-peak 
and $\alpha$-elements with a high degree of homogeneity.
}
   
	   
\titlerunning{Chemical abundances in the ONC and OB1b}
\authorrunning{K. Biazzo et al.}
\maketitle

\section{Introduction}

Measurements of elemental abundances in low-mass members of young clusters and associations represent important 
tools for addressing different issues in the field of star and planet formation.

On the one hand, elemental abundance determinations in young associations, as for old populations, allow one to use 
{\it chemical tagging} to investigate formation scenarios and possible common origin of different groups. In 
particular, accurate measurements of abundances and abundance ratios in members of different regions and subgroups 
belonging to the same star-forming complex can unveil group-to-group differences and chemical enrichment. 
In turn, this would represent a signature of sequential star formation 
and supernova nucleosynthesis (e.g., \citealt{cunhaetal98}, and references therein). 

On the other hand, giant gas planets are preferentially found around old solar-type stars more metal-rich than the Sun 
(e.g., \citealt{johnsonetal2010}, and references therein). Since planets are assumed to form from circumstellar (or proto-planetary) disks 
during the pre-main sequence (PMS) phase, the obvious question arises about the metallicity of young solar analogs and what fraction of them 
(if any) is metal-rich.

Orion is one of the most well-studied star-forming complexes and represents an ideal laboratory for investigating all the stages related 
to the birth of stars and planetary systems. In particular, the different ages of the four subgroups 
($\sim$8--12 for 1a, $\sim$3--6 for 1b, $\sim$2--6 for 1c, and $\ltsim$1--3 for 1d; \citealt{Bally2008}) belonging 
to the OB1 association, along with the content of dust and gas, appear to support the idea of sequential star-formation scenario 
(\citealt{blaauw64}), with the 1d subgroup (the Orion Nebula Cluster - ONC) being the youngest. As discussed in 
detail by \cite{cunhalamb92}, since massive stars are a major site of nucleosynthesis, the gas from which the 
younger subgroups formed as a second generation may be contaminated by the enriched ejecta of the first generation 
of massive stars (\citealt{reeves72, reeves78, preibzinne2006}). In this case, one would expect to detect different 
abundance pattern across the cluster subgroups, with the youngest regions - the ONC in particular - showing 
peculiar chemical enrichment in iron-peak and $\alpha$-elements with respect to the older ones. Owing to the 
large number of supernovae expected to have occurred in Orion, this prediction can be tested by accurate and 
homogeneous abundance measurements in the different subgroups. 

Moreover, the high frequency of proto-planetary disks around low-mass members of the ONC suggests that planetary systems might be 
forming around a fraction of these stars. Determining the metallicity of the cluster is important to investigating the connection 
between the early phases of planet formation and planets that formed some Gyr ago and are currently detected around 
old solar-type stars. We mention that none of the star-forming regions (SFRs) for which a metallicity is available 
is metal-rich (e.g., \citealt{santosetal2008,gonzalez-hernandez2008,dorazietal09,dorazietal10}), which is indeed 
puzzling. 

Several studies have been carried out in the past two decades designed to measure the abundances of the gas and 
stars in Orion. As summarized by \cite{dorazietal09}, however, these studies have yielded discrepant results 
in term of both the average metallicities of the different subgroups, the ONC in particular, and the presence 
of group-to-group differences. For example, \cite{cunhalamb94} detected variations in oxygen and silicon abundances 
across Orion, which they interpreted as the signature of self-enrichment. However, \cite{simondiaz2010} analyzed 
high quality spectra of 13 B-type stars in Orion OB1a,b,c,d and found a high degree 
of homogeneity, in contrast to the results of Cunha and collaborators.

The most recent determination of [Fe/H] in Orion based on late-type stars was performed by \cite{dorazietal09}, 
who analyzed a small sample of cool ONC members and one candidate member of the OB1b association. 
They inferred a solar metallicity for the ONC with a very small star-to-star dispersion ([Fe/H]=$-0.01\pm 0.04$), along with hints 
of a slightly sub-solar metallicity for OB1b. While this result might provide support to the sequential star formation 
scenario, \cite{dorazietal09} emphasized that their results should be confirmed based on a larger sample of stars 
and the analysis of spectra more suitable for abundance measurements.

Here, we present a new study of the elemental abundances in Orion, based on high-resolution spectra obtained with {\sc flames/uves} on the 
Very Large Telescope (VLT). Not only is our sample larger than that of \cite{dorazietal09}, but the spectral range covered by our 
spectra includes a significantly larger number of \ion{Fe}{i} and \ion{Fe}{ii} lines, which enabled us to achieve 
a more secure determination of stellar parameters. Furthermore, for the coolest stars in the sample, the analysis 
was performed using GAIA models (\citealt{hauschildt1999,brotthauschildt10}), which are more appropriate than 
ATLAS models (\citealt{Kuru93}), because of the inclusion of millions of molecular lines in the line list, as 
explained in detail in Appendix \ref{appendix:b}. Finally, the sample of \cite{dorazietal09} was reanalyzed.

In Sect.~\ref{sec:sample}, we describe the sample, observations, and data reduction. The measurements of radial 
velocity, effective temperature, veiling, and elemental abundance are given in Sects.~\ref{sec:rad_vel} and 
\ref{sec:abundance}. The results, discussion, and conclusions are presented in Sects.~\ref{sec:results}, 
\ref{sec:discussion}, and \ref{sec:conclusion}. In Appendix~\ref{appendix:a}, we give the line list, while in 
Appendix \ref{appendix:b} we describe the impact of model atmospheres on abundance measurements.

\section{Sample, observations, and data reduction}
\label{sec:sample}
\subsection{The sample}

The ONC target stars were taken from \cite{hillenbrand97}; we selected stars with spectral types from late-G 
to early-M without evidence of strong accretion and, thus, spectral veiling. We avoided stars with large 
rotational velocities ($v\sin i>30$ km/s) and known to be binaries. The total sample contains 10 stars. 
Similar criteria were applied to the OB1b group, where we selected 10 late-K stars from the 
\cite{bricenoetal05,bricenoetal07} samples.

The ONC and OB1b samples are listed in Table~\ref{tab:literature}, along with information from the literature. 
For the ONC, we indicate in Cols. 1-7 the star name, $I$ magnitude, $V-I$ color, spectral type, effective 
temperature, luminosity, and membership probability from \cite{hillenbrand97} and \cite{hillenbrandetal98}, 
and in Cols. 8-9 the values of $v sin i$ (\citealt{wolff2004,sicilia05,santosetal2008}) and some notes. 
For OB1b, we list the star name, $V$ magnitude, $V-I$ color, spectral type, effective temperature from spectral-type 
using the \cite{kenyonhartmann95} scale, luminosity, and object class taken from \cite{bricenoetal05}. In the 
last two columns of the table, we report our radial velocity measurements and comments on membership (see Sect.~\ref{sec:rad_vel}).

\begin{figure*}	%[b!]
\begin{center}
\begin{tabular}{c}
\includegraphics[width=13cm]{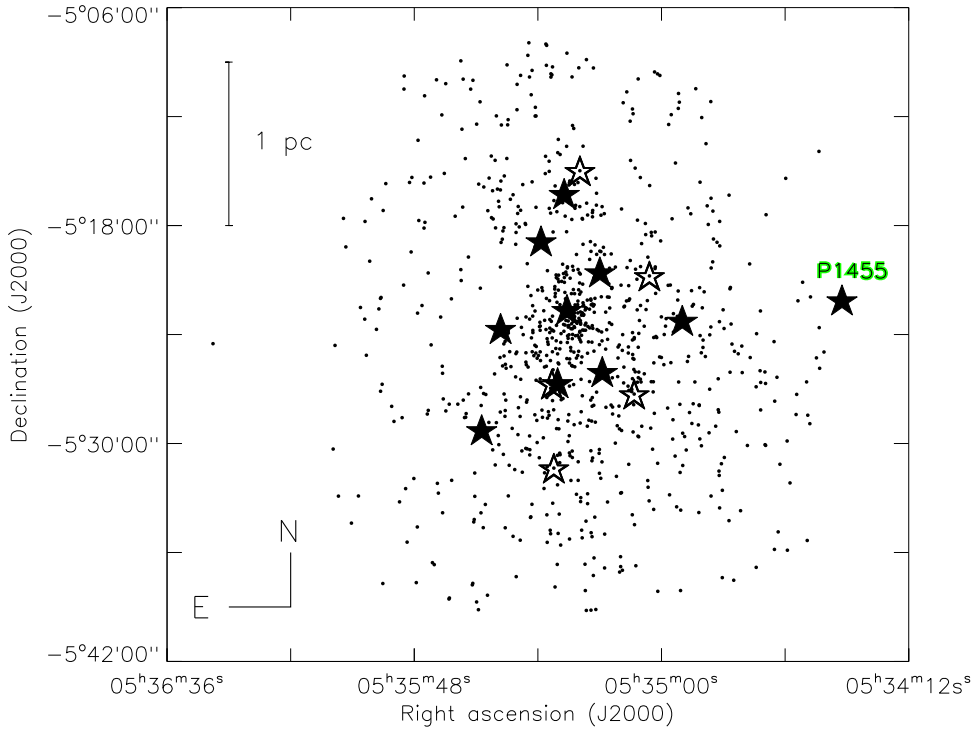}
\end{tabular}
\vspace{-.9cm}
       \caption{Spatial distribution of our ONC targets (filled stars) and the sample of \cite{dorazietal09} re-analyzed in this work (empty stars). 
       Dots represent the \cite{hillenbrand97} sample with membership probability higher than $90\%$. The field is centered on the Trapezium cluster 
       and covers an area of about 0.5\degr$\times$0.5\degr. The position of the star at the edge of the main cluster (P1455) is given.}
       \label{fig:2MASS_targets_ONC}
 \end{center}
\end{figure*}

\begin{figure*}	%[h]
\begin{center}
 \begin{tabular}{c}
\includegraphics[width=17cm]{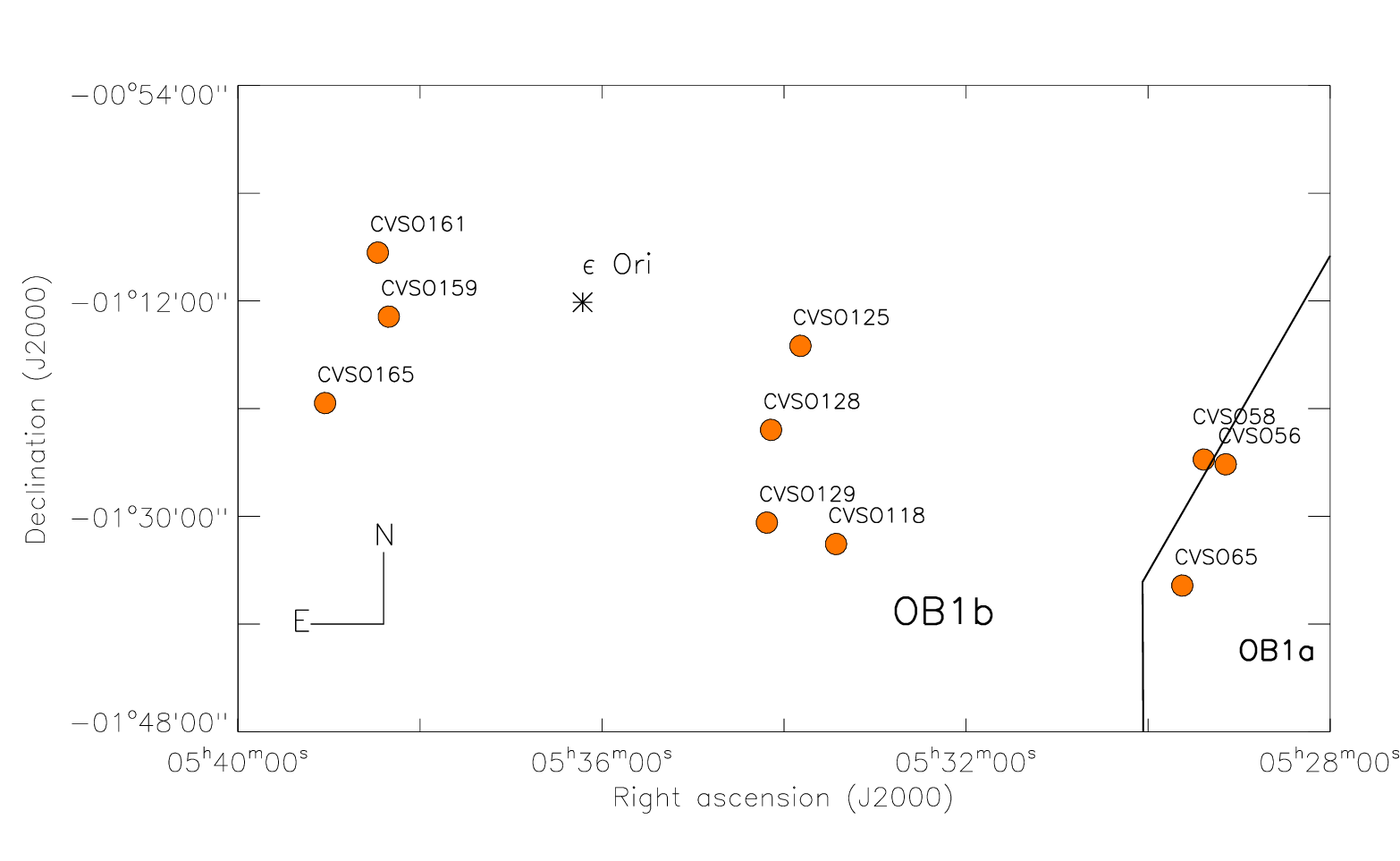}
 \end{tabular}
\vspace{-.7cm}
       \caption{Spatial distribution of our OB1b stars (dots). The field covers an area of $\sim$3\degr$\times$1\degr. The position 
       of $\epsilon$~Ori (one of the Orion Belt stars) is also shown. The solid line outlines the boundary between Orion 
       OB1a and OB1b (\citealt{warrenhesser1977}).}
       \label{fig:2MASS_targets_OB1b}
 \end{center}
\end{figure*}

In Figs.~\ref{fig:2MASS_targets_ONC} and \ref{fig:2MASS_targets_OB1b}, we show the distribution in the sky of 
our targets. The ONC stars fall inside the main cluster, with the exception of P1455. The case of this 
star is discussed in Sect.~\ref{sec:p1455}. As for the OB1b targets, three of them fall 
close to the OB1b/OB1a boundary defined by \cite{warrenhesser1977} and is discussed in Sect.~\ref{sec:membership}. 

\begin{sidewaystable*}
%\setlength{\tabcolsep}{1.pt}
%\begin{table*}[b]  
\caption{Sample stars. For the ONC stars we list: object name, $I$ magnitude, $V-I$ color, spectral type, 
photometric effective temperature, luminosity, membership probability, $v\sin i$, and notes. For the OB1b star we give: 
star name, $V$ magnitude, $V-I$ color, spectral type, effective temperature from spectral type, luminosity, 
and object class. In the last two columns our radial velocity values and comments are listed for both regions.}
\label{tab:literature}
%\tiny
\begin{center}  
\begin{tabular}{lccrrcccccc}
\hline
\hline
~\\
\multicolumn{11}{c}{ONC}\\
\hline
Star$^{a, c}$ & $I$ & $V-I$ & Sp. Type& $T_{\rm eff}^{\rm P}$ & $\log (L/L_{\odot})$ & Mem& $v\sin i$& Notes$^{d}$& $V_{\rm rad}$& Comment$^{e}$\\
              & (mag) &     &         &  (K)    	      &               & (\%) & (km/s) & & (km/s)&  \\	 
\hline
JW365 & 11.70 & 1.96 & K2-3	      & 4775 & 0.82& 99& 67.6$\pm$5.8 & NIR, WTTS:        & ... &  SB2\\    
JW373 & 13.48 & 3.37 & K2-7	      & 4775 & 0.99& 99& ...	       & HH, NIR, CTTS    & ... &   SB1\\    
JW589 & 11.30 & 2.12 & G8-M0	      & 5236 & 1.23& 99& 14	       & r	          &  27.1$\pm$0.8& M\\    
JW601 & 12.52 & 1.25 & K2-5	      & 4775 & 0.05& 98& 12.3$\pm$0.9 & WTTS	          &  22.0$\pm$0.4& M\\   
JW641 & 11.51 & 1.74 & mid-G/early-K  & 5236 & 0.91& 99& 46.3$\pm$4.1  &                  &  ...   &  SB2\\ 
JW733 & 13.14 & 2.33 & M0.5 	      & 3724 & 0.04& 99& ... &	                          & 27.4$\pm$0.1 & M\\    
JW868 & 13.19 & 1.96 & K3-5 	      & 4581 & 0.16& 99& 11.7$\pm$1.3	       &NIR, HH   & 24.7$\pm$0.5 & M\\ 
JW907 & 12.77 & 1.60 & K1-4 	      & 4775 & 0.17& 99& 32.8$\pm$2.2	       &WTTS	  & 22.9$\pm$0.2 & M\\	 
JW157 & 10.15 & 1.56 & K0-4 	      & 4775 & 1.19& 99& 7		       &N	  & 28.0$\pm$0.1 & M\\
P1455 & 10.08 & 0.76 & G0-1 	      & 5902 & 1.06& 97& 21		       &          & 21.2$\pm$0.2 & M\\	  
\hline
~\\
\multicolumn{9}{c}{OB1b}\\
\hline
Star$^{b, c}$ & $V$ & $V-I$ & Sp. Type& $T_{\rm eff}^{\rm ST}$  & $\log (L/L_{\odot})$  & Notes$^{f}$& $V_{\rm rad}$&Comment$^{e}$ \\
& (mag) &    &  & (K)& &  & (km/s)&  \\	 
\hline
CVSO118 & 14.70 & 1.26 &  K5 & 4350 & $-$0.43&   WTTS & 32.1$\pm$0.4 & M\\
CVSO125 & 14.53 & 1.34 &  K5 & 4350 & $-$0.33&   WTTS & 27.3$\pm$1.4 &   M\\
CVSO128 & 14.65 & 1.50 &  K6 & 4205 & $-$0.17&   WTTS & ...&    SB1   \\ 
CVSO129 & 14.15 & 1.23 &  K6 & 4205 & $-$0.17&   WTTS & ...&    SB1  \\ 
CVSO159 & 14.83 & 1.85 &  K7 & 4060 & $-$0.14&   WTTS & 26.0$\pm$1.3 &   M\\ 
CVSO161 & 14.32 & 1.40 &  K6 & 4205 & $-$0.19&   WTTS & 31.7$\pm$0.9 &   M\\ 
CVSO165 & 13.73 & 1.40 &  K6 & 4205 &	 0.15&   CTTS & 31.4$\pm$0.7 &   M\\ 
CVSO56  & 14.81 & 1.53 &  K7 & 4060 & $-$0.33&   WTTS & 18.7$\pm$0.3 & PM  \\ 
CVSO58  & 14.79 & 1.52 &  K7 & 4060 & $-$0.24&   CTTS & 18.7$\pm$1.0 & PM  \\ 
CVSO65  & 14.47 & 1.35 &  K6 & 4205 & $-$0.33&   WTTS & 10.2$\pm$0.3 & NM  \\ 
\hline
\end{tabular}
\end{center}
$^{a}$: JW=\cite{joneswalker88}, P=\cite{parenago1954}; $^{b}$: CVSO=CIDA Variability Survey of Orion (\citealt{bricenoetal05}).\\
$^{c}$: 2MASS (\citealt{Cutrietal2003}) data for all the ONC/OB1b targets; {\it Spitzer} (\citealt{Rebulletal2006}) data at $3.6<\lambda<8~\mu$m for JW373 and JW157; 
Infrared (\citealt{Gezarietal1999}) data at $1.25<\lambda<25~\mu$m for JW157, JW373, JW589, and JW601\\
$^{d}$: HH=Herbig-Haro object of host star; r=radio continuum source; N=N-band excess; NIR=photometric near-infrared excess. Notes from \cite{Feigelson02} 
and \cite{sicilia05}.\\
$^{e}$ SB=Spectroscopic binary (1: single-lined; 2: double-lined); M=Member; PM=Probable OB1a/25~Ori member; 
NM=non-member of OB1b.\\
$^{f}$: WTTS: Weak-lined T-Tauri star; CTTS: Classical T-Tauri star. Notes from \cite{bricenoetal05}.
%\normalsize
%\end{table*}  
\end{sidewaystable*}

\subsection{Observations and data reduction} 
The observations were obtained in 2009 with the Fiber Large Array Multi-Element Spectrograph ({\sc flames}; \citealt{pasquinietal02}) 
attached to the Kueyen Telescope (UT2) at Paranal Observatory. Both the ONC and OB1b were observed using the fiber link to {\sc uves}. 
We allocated fibers to a maximum of seven stars, leaving at least one fiber for the sky acquisition. We used the CD\#3 cross-disperser 
covering the range 4770--6820 \AA~at the resolution $R=47\,000$. This setup allowed us to select around 60+9 \ion{Fe}{i}+\ion{Fe}{ii} lines, 
as well as spectral features of $\alpha$- and Fe-peak elements (see Sect. \ref{sec:lin_sol_ew}). 

The ONC was covered with two different pointings, each including seven stars with an overlap of four stars. 
For each pointings, we obtained three 45 min long exposures, resulting in a total integration time of 2 h and 
15 min. Three pointings were instead necessary to observe 
the OB1b stars. The pointings included four, three, and three stars. Each field was observed six, two, and four times, for a total exposure 
time of 4.5, 1.5, and 3 h. The log book of the observations is given in Table~\ref{tab:observations}.

Data reduction was performed using the {\sc flames/uves} pipeline (\citealt{modiglianietal2004}) and 
the following procedure: subtraction of a master bias, order definition, extraction of thorium-argon spectra, 
normalization of a master flat-field, extraction of the science frame, wavelength calibration of the science frame, and correction 
of the science frame for the normalized master flat-field. Sky subtraction was performed with the task {\sc sarith} in the 
IRAF\footnote{IRAF is distributed by the National Optical Astronomy Observatory, which is operated by the Association of the 
Universities for Research in Astronomy, inc. (AURA) under cooperative agreement with the National Science Foundation.} {\sc echelle} 
package using the fibers allocated to the sky.

\begin{figure*}	%[b!]
\begin{center}
 \begin{tabular}{c}
  \includegraphics[width=15cm]{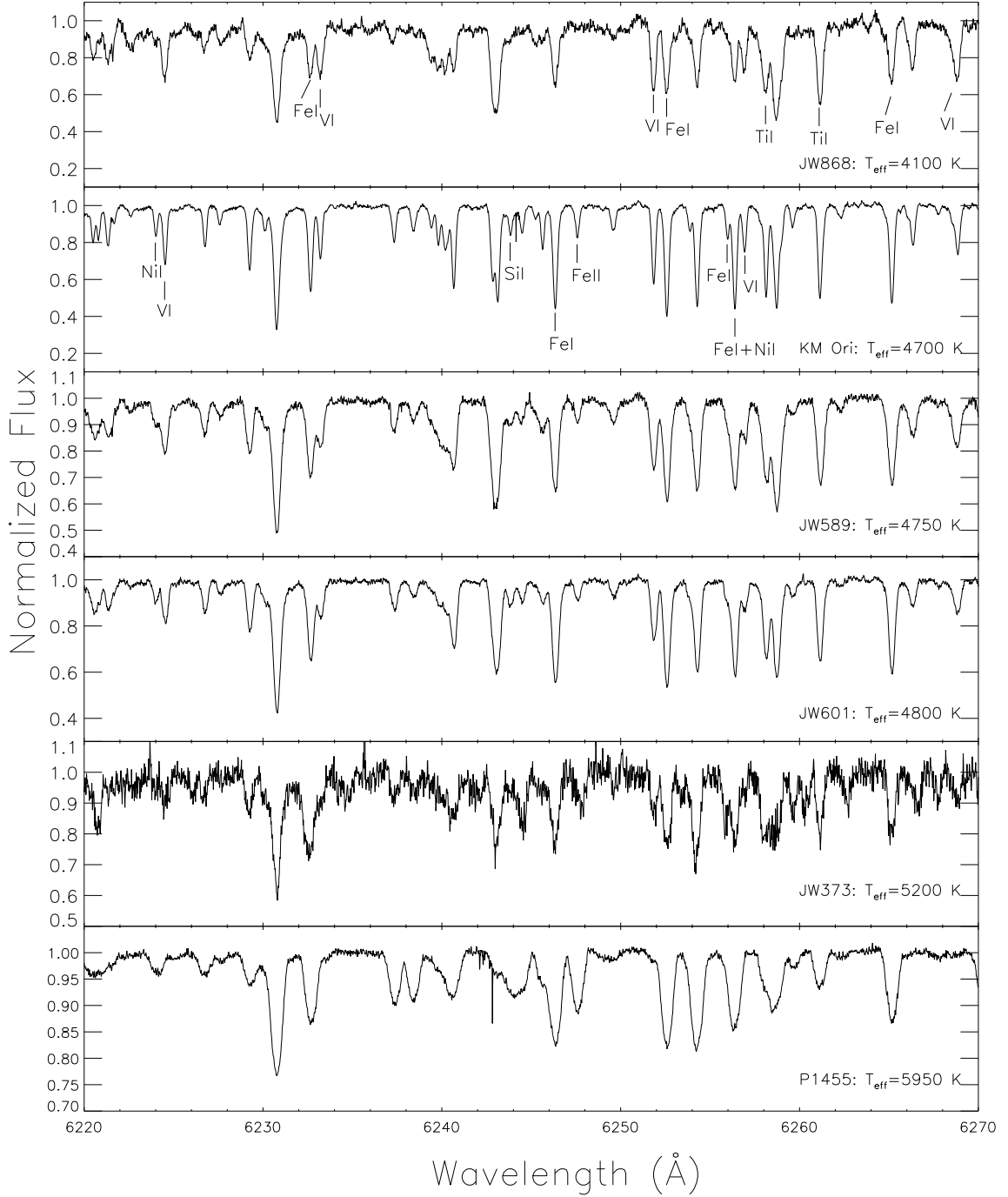}\\                  
 \end{tabular}
       \caption{Portion of spectra in the 6220-6270 \AA~wavelength range of the ONC stars for which we measured 
       the abundances. Different features used for temperature determination using line-depth ratios and 
       abundance measurements are indicated.}
       \label{fig:spectra_ONC}
 \end{center}
\end{figure*}

\begin{figure*}	%[b!]
\begin{center}
 \begin{tabular}{c}
  \includegraphics[width=15cm]{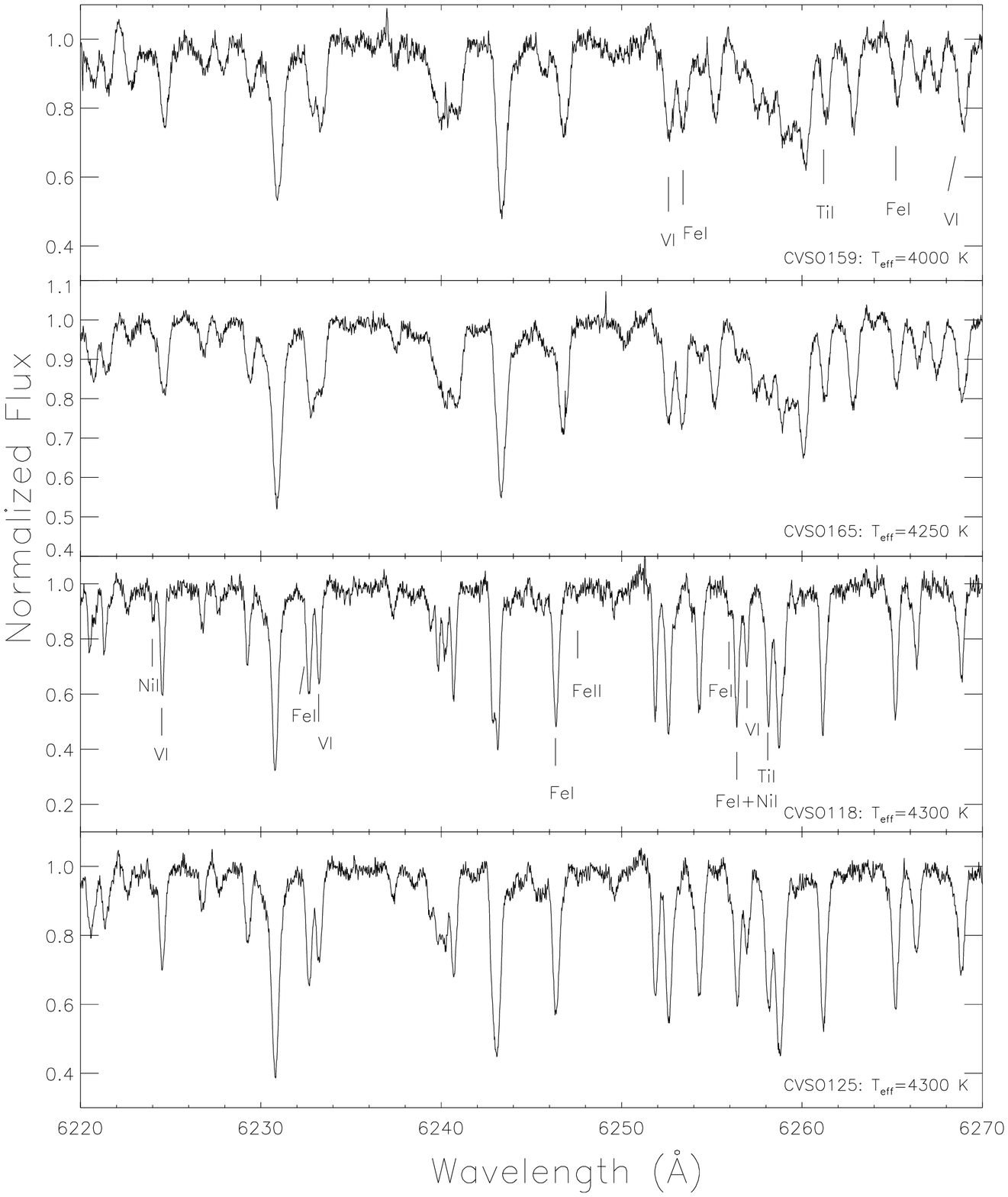}\\                  
 \end{tabular}
       \caption{As in Fig.~\ref{fig:spectra_ONC} but for the OB1b sample.}
       \label{fig:spectra_OB1b}
 \end{center}
\end{figure*}

All the acquired spectra for each star were shifted in wavelength for the heliocentric correction and then coadded, 
after checking for possible radial velocity variations. The final signal-to-noise ratio ($S/N$) is in the range 
40--200 for the ONC stars and 30--80 for the fainter OB1b targets. The coadded spectra of the stars we used 
for abundance measurements are shown in Figs.~\ref{fig:spectra_ONC} and \ref{fig:spectra_OB1b}.

\setlength{\tabcolsep}{4.5pt}
\begin{table}[h]  
\caption{Log of the observations.}
\label{tab:observations}
\begin{center}  
\begin{tabular}{cccccc}
\hline
\hline
$\alpha$ (J2000)   & $\delta$ (J2000)  &  Date      &  UT      & $t_{\rm exp}$ & $\#$\\
(h:m:s)     &  (\degr: $^\prime$ : \arcsec)  & (d/m/y)    & (h:m:s)  &  (s)          & (stars) \\ 
\hline
~\\
\multicolumn{6}{c}{ONC}\\
\hline
05:35:12 & $-$05:23:00 & 23/01/2009 & 00:56:05 & 2775 & 7 \\ 
05:35:12 & $-$05:23:00 & 23/01/2009 & 01:44:32 & 2775 & 7 \\ 
05:35:12 & $-$05:23:00 & 23/01/2009 & 02:45:16 & 2775 & 7 \\ 
05:35:16 & $-$05:24:20 & 28/01/2009 & 02:04:19 & 2775 & 7 \\ 
05:35:16 & $-$05:24:20 & 30/01/2009 & 01:56:17 & 2775 & 7 \\ 
05:35:16 & $-$05:24:20 & 30/01/2009 & 02:44:12 & 2775 & 7 \\ 
\hline
~\\
\multicolumn{6}{c}{OB1b}\\
\hline
05:29:27 & $-$01:31:60 & 22/01/2009 & 00:54:00 & 2775 & 3\\
05:29:27 & $-$01:31:60 & 31/01/2009 & 00:59:31 & 2775 & 3\\
05:29:27 & $-$01:31:60 & 31/01/2009 & 01:47:38 & 2775 & 3\\
05:29:27 & $-$01:31:60 & 01/02/2009 & 00:53:31 & 2775 & 3\\
05:33:48 & $-$01:25:59 & 28/01/2009 & 01:02:48 & 2775 & 4\\
05:33:48 & $-$01:25:59 & 14/02/2009 & 01:41:02 & 2775 & 4\\
05:33:48 & $-$01:25:59 & 18/02/2009 & 00:34:47 & 2775 & 4\\
05:33:48 & $-$01:25:59 & 18/02/2009 & 01:22:50 & 2775 & 4\\
05:33:48 & $-$01:25:59 & 24/02/2009 & 00:31:30 & 2775 & 4\\
05:33:48 & $-$01:25:59 & 24/02/2009 & 01:32:14 & 2775 & 4\\
05:38:35 & $-$01:11:60 & 16/01/2009 & 01:20:37 & 2775 & 3\\
05:38:35 & $-$01:11:60 & 21/01/2009 & 02:37:49 & 2775 & 3\\
\hline
\end{tabular}
\end{center}
\end{table}  

\section{Radial velocities and membership}
\label{sec:rad_vel}
Since we did not acquire spectra of radial velocity (RV) standards, we first selected two stars to be used as templates in the ONC and OB1b: 
JW601 (ONC) and CVSO118 (OB1b). Both have high $S/N$ spectra ($\sim$50--100), low $v\sin i$ ($<$15 km s$^{-1}$), and radial velocity 
measurements from the literature. In particular, CVSO118 is the only star in our OB1b sample with a previous determination of RV. 
We then measured the RV from the first spectrum we acquired for both stars using the IRAF task {\sc rvidlines} inside the {\sc rv} 
package. This task measures RVs from a line list and we used 25 lines in the spectral range 5800--6800 \AA. For JW601, we obtain 
$V_{\rm rad}=$22.8$\pm$0.5 km s$^{-1}$, while for CVSO118 we find that $V_{\rm rad}=$32.1$\pm$0.4 km s$^{-1}$. These values are in very good agreement with 
previous determinations of 22.0$\pm$0.4 km s$^{-1}$ and 31.6$\pm$0.6 km s$^{-1}$ obtained by \cite{sicilia05} and \cite{bricenoetal07}, 
respectively.

Considering JW601 and CVSO118 as templates, we measured the heliocentric RV of all the ONC/OB1b stars using the task {\sc fxcor} of 
the IRAF package {\sc rv}. To take advantage of the wide spectral coverage offered by {\sc flames/uves}, we cross-correlated 
all the spectral range of our targets with the template, excluding the regions contaminated by broad emission lines (e.g., H$\alpha$) 
or by prominent telluric features (e.g., the O$_2$ series at $\lambda\simeq6275$\,\AA). To determine in the most 
reliable way the centroids of the cross-correlation function (CCF) peaks, we adopted Gaussian fits. The errors in 
the RV values were computed using a procedure inside the {\sc fxcor} task that considers the fitted peak height 
and the antisymmetric noise as described by \cite{TonryDavis1979}. 
Since we acquired several spectra per stars, we computed an average RV value for all our targets that did not 
display evidence of binarity. The average RV values for most probably single stars are listed in 
Table~\ref{tab:literature}.

\subsection{Membership}
\label{sec:membership}
\begin{figure}	%[b!]
\begin{center}
 \begin{tabular}{c}
  \resizebox{\hsize}{!}{\includegraphics{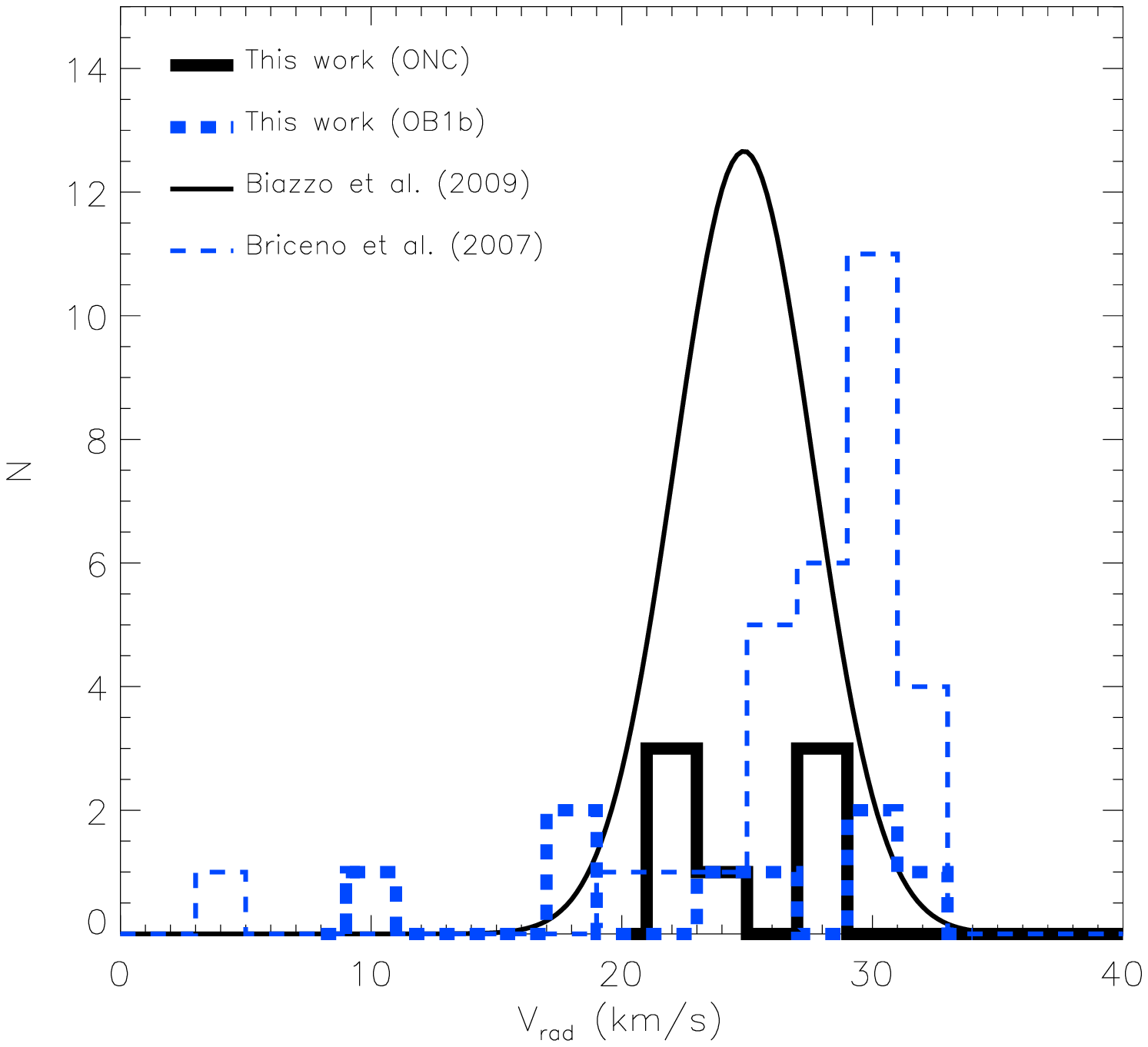}}\\                  
 \end{tabular}
       \caption{Radial velocity distribution for the most probable single stars in the ONC (solid lines) and OB1b (dashed lines) samples. 
       The thick solid and dashed lines refer to this work. The thin solid line represents the Gaussian fit to the \cite{biazzoetal2009} 
       sample obtained for 96 ONC targets with a mean $V_{\rm rad}=$24.87 km s$^{-1}$ ($\sigma_V{}_{\rm rad}=$2.74 km s$^{-1}$). 
       The thin dashed line represents the histogram obtained by \cite{bricenoetal07} for 30 OB1b stars with mean 
       $V_{\rm rad}=30.1\pm1.9$ km s$^{-1}$.}
       \label{fig:vrad_distr}
 \end{center}
\end{figure}

To confirm the membership of our stars, the distributions of average RV measurements for the ONC are shown 
in Fig.~\ref{fig:vrad_distr}, along with the value derived by \cite{biazzoetal2009} for $\sim$100 very low-mass 
members. Seven out of 10 stars of ONC are confirmed as members and most probably single stars with a mean RV 
of 24.9$\pm$2.6 km s$^{-1}$ in good agreement with previous determinations (\citealt{sicilia05, biazzoetal2009}). 
The three exceptions are JW373, JW365, and JW641, which exhibit a double/triple-peaked CCF. We, thus, classify 
these stars as, previously unidentified, binaries. Two of them (JW365 and JW641) also have a double-lined system 
(SB2) and are therefore discarded from further analysis.

For OBab, similarly, we compare our sample with the distribution found by \cite{bricenoetal07} based on 30 OB1b 
targets from which they derived a mean RV of $30.1\pm1.9$ km s$^{-1}$. Five sample stars are confirmed single 
members, with an average RV of 29.9$\pm$3.0 km s$^{-1}$, while three are non-members and two (CVSO129 and CVSO128) 
have double/triple-peaked CCFs. Among the non-members, CVSO56 and CVSO58 have RVs close to the OB1a/25~Ori region 
(mean RV $\sim$20 km s$^{-1}$; \citealt{bricenoetal05}), as also implied by their spatial location close to the 
OB1b/OB1a boundary (Fig.~\ref{fig:2MASS_targets_OB1b}). 

\section{Abundance analysis}
\label{sec:abundance}
To summarize, abundances were obtained for all the ONC and OB1b single members or single-lined binaries, with 
the exception of rapid rotators (JW907, CVSO128, CVSO129, and CVSO161) and the very cool star JW733 for which 
measuring abundances from line equivalent widths (EW) is not suitable. 
Since CVSO56 seems to belong to the OB1a/25~Ori subgroup, we measured its iron abundance to gauge the properties of this region. 
On the other hand, we have not been able to derive the metallicity of CVSO58 because of its rapid rotation.

The analysis was performed using the 2002 version of MOOG (\citealt{sneden1973}) that assumes local thermodynamic equilibrium (LTE) 
and where the radiative and Stark broadening are treated in a standard way. For collisional broadening, we used the \cite{unsold1955} approximation. 
Both \cite{Kuru93} and \cite{brotthauschildt10} grids of plane-parallel model atmospheres were used for stars 
warmer and cooler than $\sim 4400$ K, 
respectively (see Appendix~\ref{appendix:b}). This is the major change introduced by ourselves with respect to 
the study of \cite{dorazietal09}. 

\subsection{Spectral veiling}
\label{sec:veiling}
We estimated the amount of veiling that affects the spectra of our stars following the procedure described 
by \cite{dorazietal09}. 
In particular, we selected the nine lines in their list included in our spectral range. We then compared 
the equivalent widths of these lines to those measured in the spectra of 16 members of the open clusters 
IC~2602 and IC~2391, which are old enough (30-50 Myr; \citealt{Randichetal2001}) to ensure that their spectra are not affected by veiling. 
These IC stars have effective temperatures similar to the ONC/OB1b targets ($\sim$4300-5800 K) and their spectra 
have a resolution close to ours.
For each line, we checked whether any dependence of the EW on effective temperature was present; since for most lines 
we found a weak trend within an interval of 1000 K, we decided to bin the whole temperature range in 500 K steps 
and derive the mean EWs of the IC stars inside each bin. For two lines, namely \ion{Ca}{i} 6102.7 \AA\,and 
\ion{Ca}{i} 6122.2 \AA, we found significant trends at all temperatures, hence we used the EW value of the 
IC cluster star with temperature closer to that of our ONC/OB1b star. For each line, we then obtained the 
veiling as $r_{\rm line}=\frac{EW_{\rm IC}}{EW_{\rm ONC/OB1b}}-1$. The mean veiling $<r>$ was then computed 
as the average of all $r_{\rm line}$ values.

By applying this method, we determined a veiling value consistent with zero for all the stars, with the exception of JW373 ($r=0.128\pm0.080$) 
in the ONC, which is indeed classified as CTTS (Table~\ref{tab:literature}). We thus corrected the measured EWs of all the lines 
using the relationship between the true EW and the measured one: $EW_{\rm true}=EW_{\rm meas}(1+<r>)$.

\subsection{Line list, solar analysis, and EWs}
\label{sec:lin_sol_ew}
We adopted the line list of \cite{randichetal2006} integrated with lines from the list of \cite{dorazirandich09} 
included in our wavelength range. We refer to both papers for details on atomic parameters and their sources. 
The line list is given in Table~\ref{tab:line_list}.

As usually done, our analysis was performed differentially with respect to the Sun. We analyzed the \cite{randichetal2006} solar spectrum 
obtained with {\sc uves}, using our combined line list and their solar parameters ($T_{\rm eff}=5770$ K, $\log g=4.44$, $\xi=1.1$ km s$^{-1}$). 
We obtained $\log n{\rm (Fe)}=7.52\pm0.02$ for \cite{Kuru93} models and $\log n{\rm (Fe)}=7.51\pm0.02$ 
for \cite{brotthauschildt10} models. The results for all the elements are given in Table~\ref{tab:solar_abun} together with 
those given by \cite{AndersGrev89} and \cite{asplundetal2009}. We caveat that the latter values were obtained 
using 3D models. Table~\ref{tab:solar_abun} shows good agreement between the two model atmospheres and between our 
and literature values.

The EWs of the target stars were measured by means of a direct integration or Gaussian fitting procedure using the IRAF {\sc splot} 
task. Very strong lines ($EW\gtsim150$ m\AA), which are most affected by the treatment of damping, were excluded from the list; furthermore, 
a 2-$\sigma$ clipping was applied to the \ion{Fe}{i} list before determining stellar parameters and iron abundance. The abundance 
of the other elements was derived using the same criteria.

\begin{table}[b]  
\caption{Comparison between solar abundances derived using \cite{Kuru93} and \cite{brotthauschildt10} model atmospheres. 
The standard values from \cite{AndersGrev89} and \cite{asplundetal2009} are also listed.}
\label{tab:solar_abun}
\begin{center}
\begin{tabular}{lcccc}
\hline
\hline
Element & $\log n_{\rm ATLAS}$ & $\log n_{\rm GAIA}$ & $\log n_{\rm AG89}$& $\log n_{\rm AS09}$\\
\hline
\ion{Na}{i} & 6.31$\pm$0.03 & 6.29$\pm$0.03 & 6.33 & 6.24$\pm$0.04\\
\ion{Al}{i} & 6.48$\pm$0.03 & 6.47$\pm$0.03 & 6.47 & 6.45$\pm$0.03\\
\ion{Si}{i} & 7.56$\pm$0.03 & 7.53$\pm$0.03 & 7.55 & 7.51$\pm$0.03\\
\ion{Ca}{i} & 6.35$\pm$0.03 & 6.34$\pm$0.03 & 6.36 & 6.34$\pm$0.04\\
\ion{Ti}{i} & 4.97$\pm$0.02 & 4.97$\pm$0.02 & 4.99 & 4.95$\pm$0.05\\
\ion{Fe}{i} & 7.52$\pm$0.02 & 7.51$\pm$0.02 & 7.52 & 7.50$\pm$0.04\\
\ion{Ni}{i} & 6.26$\pm$0.03 & 6.24$\pm$0.02 & 6.25 & 6.22$\pm$0.04\\
\hline		
\end{tabular}
\end{center}
\end{table}

\subsection{Stellar parameters}
\label{sec:parameters}

\subsubsection{Effective temperatures}

Photometric temperatures were taken from \cite{hillenbrand97} for the ONC ($T_{\rm eff}^{\rm P}$ in 
Table~\ref{tab:literature}), while for the OB1b sample we converted the spectral-types of \cite{bricenoetal05} to temperatures 
using the scale of \cite{kenyonhartmann95} ($T_{\rm eff}^{\rm ST}$ in Table~\ref{tab:literature}).

Spectroscopic temperatures were then derived in two different ways. It has been demonstrated that line-depth 
ratios (LDRs) are powerful tools for measuring the effective temperature with a precision 
as small as 10--50\,K for spectra with $S/N > 100$ (\citealt{Gray1991, kovtyukhetal06, Biazzo07}, and references therein). 
The precision of this method can be improved by averaging the results from several line pairs. To develop appropriate 
$T_{\rm eff}$-LDR calibrations, we considered the synthetic stellar library described and made available by \cite{Coelho05}. 
These spectra are sampled at 0.02\,\AA, range from the near-ultraviolet (300 nm) to the near-infrared (1.8 $\mu$m), and cover the 
following grid of parameters: 3500$\le T_{\rm eff}\le$7000 K, 0.0$\le\log g\le$5.0, $-$2.5$\le$[Fe/H]$\le$+0.5, $\alpha$-enhancement 
[$\alpha$/Fe]=0.0, 0.4 and microturbulent velocity $\xi$=1.0, 1.8, 2.5 km s$^{-1}$. 

Following the prescriptions given by \cite{Biazzo07}, we used their line list in the 6190--6280~\AA~spectral range and 
their line pairs, namely 15 (see their Tables 1 and 2; we refer to that paper for a detailed explanation and justification 
of the line list and line pairs selected). We then measured their LDRs according to the guidelines of \cite{catalanoetal2002}, 
and developed $T_{\rm eff}$-LDR calibrations after degrading the synthetic spectra to our resolution. The calibrations were 
created for 3.0$\le\log g\le$4.5, 4000$\le T_{\rm eff}\le$6500 K, and $v\sin i$=0 km s$^{-1}$, considering the synthetic 
spectra at [Fe/H]=0.0, [$\alpha$/Fe]=0.0, and $\xi$=1.0 km s$^{-1}$. 

\begin{figure}	%[b!]
\begin{center}
 \begin{tabular}{c}
  \resizebox{\hsize}{!}{\includegraphics{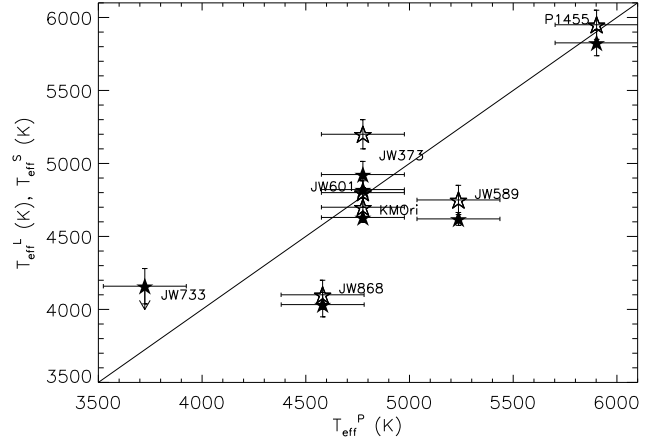}}\\                  
 \end{tabular}
       \caption{Spectroscopic effective temperatures versus the photometric values obtained by \cite{hillenbrand97}. Our 
       $T_{\rm eff}^{\rm L}$ and $T_{\rm eff}^{\rm S}$ values obtained using the LDR method and MOOG 
       code are shown as filled and empty stars, respectively. The arrow represents an upper limit. The error bars 
       on the $x$-axis refer to the values given by \cite{hillenbrand97}, while those on the $y$-axis refer to the uncertainties 
       in $T_{\rm eff}^{\rm L}$ only, while typical uncertainties in $T_{\rm eff}^{\rm S}$ are 60 K 
       (see text).}
       \label{fig:temperature_onc}
 \end{center}
\end{figure}

\begin{figure}	%[b!]
\begin{center}
 \begin{tabular}{c}
  \resizebox{\hsize}{!}{\includegraphics{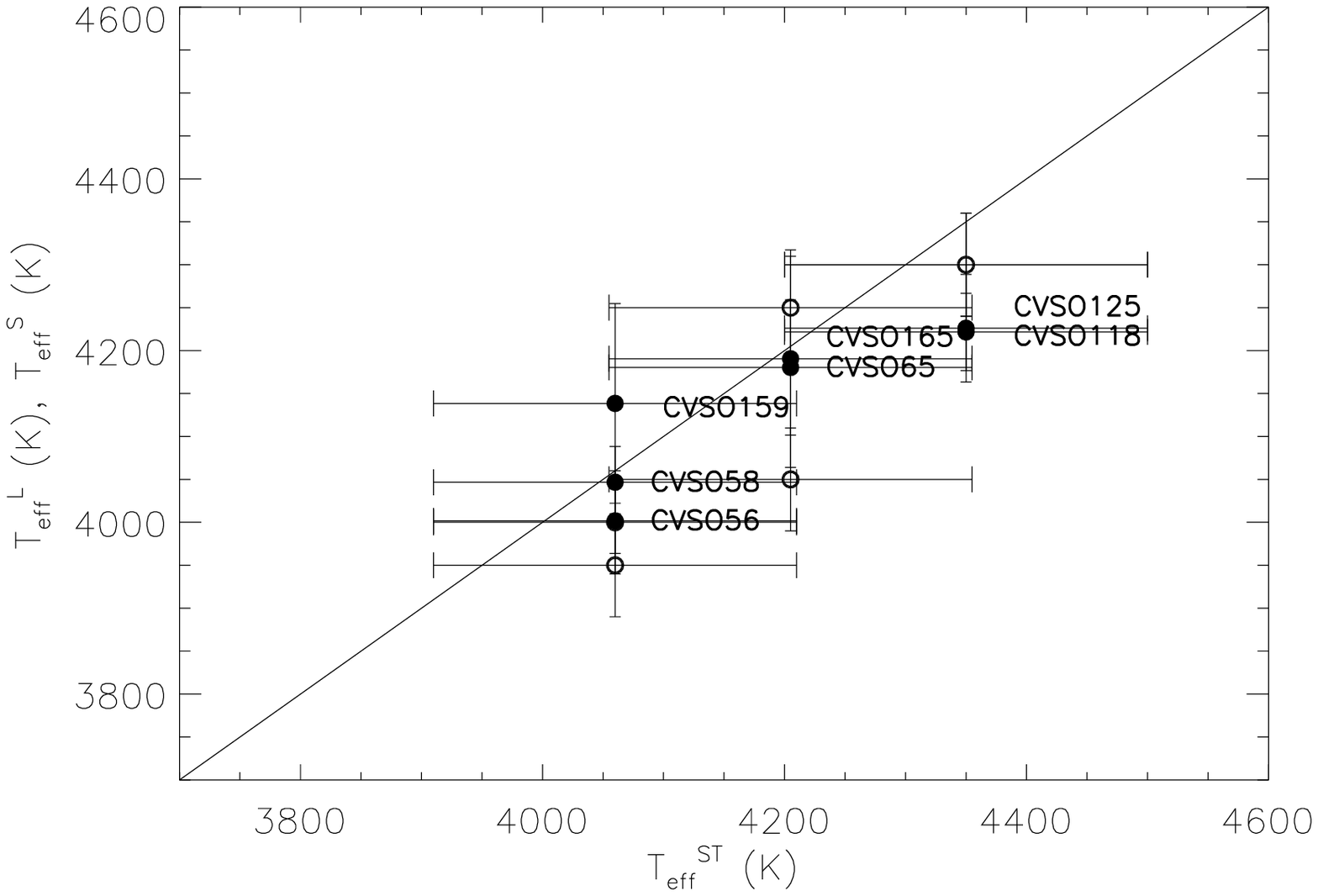}}\\                  
 \end{tabular}
       \caption{Our spectroscopic effective temperatures versus the values obtained by converting the spectral 
       types given by \cite{bricenoetal05} into temperature using the tables of \cite{kenyonhartmann95}. Our 
       $T_{\rm eff}^{\rm L}$ and $T_{\rm eff}^{\rm S}$ values obtained using the LDR method and MOOG 
       code are shown as filled and empty circles, respectively. The error bars on the $x$-axis refer to 
       the values given by \cite{bricenoetal05}, while those on $y$-axis refer to 
       the uncertainties in $T_{\rm eff}^{\rm L}$ only, while typical uncertainties in $T_{\rm eff}^{\rm S}$ are 60 K 
       (see text). Note the different temperature range from Fig.~\ref{fig:temperature_onc}.}
       \label{fig:temperature_ob}
 \end{center}
\end{figure}

In the end, the effective temperatures obtained from all the useful LDRs for each ONC/OB1b target were averaged to 
increase the precision of the temperature determination. The values ($T_{\rm eff}^{\rm L}$) are listed in Table~\ref{tab:param_chem} 
and plotted in Figs.~\ref{fig:temperature_onc} and \ref{fig:temperature_ob}.

\begin{sidewaystable*}
\setlength{\tabcolsep}{1.2pt}
%\begin{table*}[h]  
\caption{Astrophysical parameters and chemical abundances derived from our analysis.}
\label{tab:param_chem}
\begin{center}  
\tiny
\begin{tabular}{lcccccrrrrrrrrrr}
\hline
\hline
\tiny{Star}&$\tiny{T_{\rm eff}^{\rm L}}$&$\tiny{T_{\rm eff}^{\rm S}}$&$\tiny{\log g^{\rm S}}$&$\tiny{\log g^{\rm P}}$&$\xi$&\tiny{[\ion{Fe}{i}/H]}&\tiny{[\ion{Fe}{ii}/H]}
&\tiny{[Na/Fe]}&\tiny{[Al/Fe]}&\tiny{[Si/Fe]}&\tiny{[Ca/Fe]}&\tiny{[\ion{Ti}{i}/Fe]}
&\tiny{[\ion{Ti}{ii}/Fe]}&\tiny{$<$[Ti/Fe]$>$}&\tiny{[Ni/Fe]}\\ 
   &\tiny{(K)} & \tiny{(K)} & & & \tiny{(km/s)} & & & & & & & & & & \\  
\hline
~\\
\multicolumn{15}{c}{ONC}\\
\hline
JW373 & 4924$\pm$90  & 5200 & 3.2 & 3.2 & 1.8  &$-$0.15$\pm$0.17(18)&$-$0.18$\pm$0.25(4)&   0.18$\pm$0.20&   0.13$\pm$0.23&0.10$\pm$0.20(1)&	0.04$\pm$0.18(3)&    0.00$\pm$0.17(2)&  ...	    & ...&$-$0.07$\pm$0.24(3) \\   
JW589 & 4620$\pm$44  & 4750 & 3.5 & 3.5 & 1.9  &$-$0.16$\pm$0.01(29)&$-$0.16$\pm$0.07(3)&   0.17$\pm$0.10&   0.21$\pm$0.20&0.14$\pm$0.11(3)&	0.02$\pm$0.09(2)&    0.00$\pm$0.10(5)&   0.14$\pm$0.09 & 0.07$\pm$0.13&$-$0.08$\pm$0.13(9) \\       
JW601 & 4821$\pm$60  & 4800 & 4.0 & 4.1 & 2.3  &$-$0.14$\pm$0.10(38)&$-$0.12$\pm$0.13(5)&   0.11$\pm$0.09&   0.14$\pm$0.07&0.14$\pm$0.12(5)&$-$0.02$\pm$0.08(2)& $-$0.08$\pm$0.08(7)&   0.18$\pm$0.07 & 0.05$\pm$0.11&$-$0.03$\pm$0.07(13)\\	 
JW733 &$<$4159$\pm$121&	 ... & ...	& 3.1 &  ...    &  ...		    &	...		&...		    &	...		&	...	 &...		     &  ...		  &...		   &...     &	...	     \\   
JW868 & 4033$\pm$84  & 4100 &...	& 3.8 & 1.0  &$-$0.15$\pm$0.20(32)&	...		&$-$0.14$\pm$0.23&$-$0.03$\pm$0.24&0.15$\pm$0.36(2)&$-$0.19$\pm$0.22(3)&$-$0.22$\pm$0.21(4)&	0.40$\pm$0.21 & 0.09$\pm$0.29&   0.03$\pm$0.23(7) \\
JW157 & 4630$\pm$33  & 4700 & 3.1 & 3.0 & 1.9  &$-$0.15$\pm$0.08(40)&$-$0.12$\pm$0.05(7)&   0.12$\pm$0.09&   0.12$\pm$0.14&0.14$\pm$0.13(4)&   0.02$\pm$0.13(4)& $-$0.05$\pm$0.11(8)&   0.20$\pm$0.10 &0.08$\pm$0.15& $-$0.10$\pm$0.13(19)\\		
P1455& 5826$\pm$88  & 5950 & 4.1 & 3.8 & 1.8  &       0.11$\pm$0.09(32)&   0.11$\pm$0.07(5)&   0.03$\pm$0.10&   0.08$\pm$0.11&0.01$\pm$0.13(2)&   0.00$\pm$0.13(3)&    0.02$\pm$0.13(6)&$-$0.13$\pm$0.10 &$-$0.06$\pm$0.16 &$-$0.10$\pm$0.13(12)\\    
~\\
ONC$^{a}$&	 &	& &   &      &$-$0.11$\pm$0.11    & $-$0.09$\pm$0.12  &   0.12$\pm$0.06   &   0.14$\pm$0.05   &0.11$\pm$0.05   &   0.01$\pm$0.02   & 	&   	& 0.05$\pm$0.06&$-$0.06$\pm$0.05\\   
ONC$^{b}$&	 &	& &   &      &$-$0.15$\pm$0.01    & $-$0.15$\pm$0.03  &   0.15$\pm$0.04   &   0.15$\pm$0.04   &0.13$\pm$0.02   &   0.02$\pm$0.03   & 	&   	& 0.07$\pm$0.01&$-$0.05$\pm$0.05\\   
ONC$^{c}$&	 &	& &   &      &$-$0.11$\pm$0.08    &   &      &      &	&      &    &	    && \\  
ONC$^{d}$&	 &	& &   &      &$-$0.13$\pm$0.03    &   &      &      &	&      &    &	    && \\  
\hline
~\\
\multicolumn{15}{c}{OB1b}\\
\hline
CVSO118 & 4222$\pm$45 &  4300 & ...& 4.3& 1.5 &    0.01$\pm$0.12(39)&...    &$-$0.19$\pm$0.12&$-$0.10$\pm$0.12&0.11$\pm$0.13(2)&$-$0.19$\pm$0.16(3)&$-$0.30$\pm$0.14(6)&0.43$\pm$0.14&0.07$\pm$0.20 & $-$0.03$\pm$0.16(19) \\
CVSO125 & 4226$\pm$63 &  4300 & ...& 4.2& 1.7 & $-$0.03$\pm$0.12(42)&...   &$-$0.15$\pm$0.12&$-$0.04$\pm$0.12&0.15$\pm$0.16(2)&$-$0.16$\pm$0.12(2)&$-$0.33$\pm$0.13(6)&0.47$\pm$0.13& 0.07$\pm$0.19   & 0.03$\pm$0.15(11) \\
CVSO159 & 4139$\pm$116&  4000 & ...& 3.9& 1.9 & $-$0.12$\pm$0.23(21)&...   &$-$0.09$\pm$0.24&$-$0.09$\pm$0.23&...		  &$-$0.10$\pm$0.23(1)&$-$0.36$\pm$0.32(3)&...			&... &$-$0.01$\pm$0.23(2) \\
CVSO165 & 4191$\pm$127&  4250 & ...& 3.7& 1.4 & $-$0.06$\pm$0.06(22)&...   &$-$0.18$\pm$0.07&   0.02$\pm$0.06&...		  &$-$0.06$\pm$0.12(2)&$-$0.22$\pm$0.21(5)&...		   &  ... & 0.04$\pm$0.31(7) \\
CVSO56  & 4002$\pm$38 &  4000 & ...& 4.0& 1.2 & $-$0.08$\pm$0.15(33)&...   &	...	&...		    &...		     & ... 		 &...		     &  	 ...     & ...&... \\
CVSO58  & 4047$\pm$42 &	 ... & ...& 4.0&  ...   &...			 &...	 &	...	&...		    &...		     &  ...		 &	...	     &  	...      & ...& ...\\
CVSO65  & 4180$\pm$79 &  4100 & ...& 4.2& 1.6 &	...		 &...	 &...		&...		    &	...	     & ... 		 &	...	     &  ...	      & ...&... \\
~\\
OB1b    &              &		  &	  &	&     & $-$0.05$\pm$0.05    &	     &  		 &		     &0.12$\pm$0.01   & 		  &   &   & 0.07$\pm$0.01&0.01$\pm$0.03 \\
\hline	 
\end{tabular}
\end{center}
%\end{table*}  
$^{a}$ Mean abundances with P1455.\\
$^{b}$ Mean abundances without P1455.\\
$^{c}$ Mean abundances with P1455 and the D'Orazi et al. re-analysis (see Table~\ref{tab:re-analysis}).\\
$^{d}$ Mean abundances without P1455 and the D'Orazi et al. re-analysis (see Table~\ref{tab:re-analysis}).\\
\end{sidewaystable*}
\normalsize

As commonly done, effective temperatures were also determined by imposing the condition that the \ion{Fe}{i} abundance does 
not depend on the excitation potential of the lines. These temperatures are defined $T_{\rm eff}^{\rm S}$ 
(Table~\ref{tab:param_chem}) and represent the adopted values for the abundance analysis (Sect.~\ref{sec:results}). 

We first note that the two values $T_{\rm eff}^{\rm L}$ and $T_{\rm eff}^{\rm S}$ closely agree (see 
Figs.~\ref{fig:temperature_onc} and \ref{fig:temperature_ob}), with the only exception of JW373, which is, 
as mentioned, a probable binary affected by veiling that has a rather low $S/N$ spectrum. Moreover, in 
Fig.~\ref{fig:temperature_onc} the agreement of both spectroscopic temperatures with $T_{\rm eff}^{\rm P}$ is 
also good, with the exception of JW868 and JW589, which we find cooler than the \cite{hillenbrand97} values 
on average by $\sim$550 K. For JW733, we can give only an upper limit because $T_{\rm eff}$-LDR calibrations 
are suitable for temperatures higher than 4000 K. In Fig.~\ref{fig:temperature_ob}, the agreement with 
$T_{\rm eff}^{\rm ST}$ is also good within the errors.

\subsubsection{Microturbulence velocities and surface gravities}

The microturbulence velocity $\xi$ was determined by imposing that the \ion{Fe}{i} abundance is independent on 
the line equivalent widths. The initial microturbulence velocity was set to 1.5 km\,s$^{-1}$. The values of 
$\xi$ are listed in Table~\ref{tab:param_chem}. We note that our determinations are typically higher than 
those of \cite{dorazietal09}. We comment on this in Sect.~\ref{sec:previous_results}.

At variance with \cite{dorazietal09}, who did not have enough \ion{Fe}{ii} features in their spectral range, we were able to estimate the surface 
gravity $\log g$ by imposing the \ion{Fe}{i}/\ion{Fe}{ii} ionization equilibrium ($\log g^{\rm S}$ in Table~\ref{tab:param_chem}) for all the ONC 
stars with the exception of JW868, where no \ion{Fe}{ii} line was found. For the stars in the ONC, the initial $\log g$ was obtained from the 
relation between mass, luminosity, and temperature ($\log g=4.44+\log M +4 \log (T_{\rm eff}/5770)-\log L$, labeled as $\log g^{\rm P}$ in 
Table~\ref{tab:param_chem}) taking as astrophysical parameters the values given by \cite{hillenbrand97}. We verified that the effect on the gravity 
of considering $T_{\rm eff}^{\rm S}$ instead of the \cite{hillenbrand97} temperature is negligible for our accuracy. We note that the agreement 
between $\log g^{\rm S}$ and $\log g^{\rm P}$ is quite good, the difference being at most 0.30 dex.

On the other hand, since the stars in OB1b are cooler than those in the ONC, we did not find any \ion{Fe}{ii} 
lines, with the only exception of a couple of lines in CVSO118 and CVSO125. Thus, we decided to fix the surface 
gravity to the values obtained using the relation between $M$, $L$, and $T_{\rm eff}$, where the astrophysical 
parameters were taken from \cite{bricenoetal05}.

Finally, we remeasured [Fe/H] for the \cite{dorazietal09} stars applying the same method used here. We measured 
the line EWs and derived the atmospheric parameters and iron abundances using ATLAS and GAIA models for stars 
with effective temperatures higher and lower than 4400 K, respectively. 

\begin{table*}	%[t]
\caption{Our reanalysis of the \cite{dorazietal09} sample ({\it left}) and \cite{dorazietal09} outputs ({\it right}). 
The number of lines employed for the iron abundance measurements are enclosed in parentheses.}
\label{tab:re-analysis}
\begin{center}
\begin{tabular}{l|cccr|cccr}
\hline
\hline
Name & $T_{\rm eff}$ & $\log g$ & $\xi$ &  [Fe/H]$^{\rm our}$ & $T_{\rm eff}$ & $\log g$ & $\xi$ & [Fe/H]$^{\rm D'Orazi\,et\,al.}$ \\
      &  (K) &  &  (km/s) &   &  (K) &  &  (km/s) & \\
\hline
268   &  4300   &   3.9  &   1.8    & $-$0.16$\pm$0.13(26) &  4300   &   3.9  &   1.6 & $-$0.08$\pm$0.16(25)\\
683   &  4400   &   3.3  &   1.8    & $-$0.13$\pm$0.06(21) &  4250   &   3.3  &   1.6 &    0.00$\pm$0.11(22)\\ 
487   &  4400   &   3.9  &   1.6    & $-$0.06$\pm$0.07(23) &  4300   &   3.9  &   1.7 & $-$0.07$\pm$0.08(18)\\							  
223a  &  4550   &   3.8  &   2.0    & $-$0.10$\pm$0.09(16) &  4450   &   3.8  &   1.6 &    0.03$\pm$0.06(18)\\
673   &  4700   &   4.0  &   1.7    & $-$0.10$\pm$0.14(19) &  4700   &   4.0  &   1.5 &    0.00$\pm$0.10(18)\\
\hline
\end{tabular}
\end{center}
\end{table*}

\subsection{Errors}
\label{sec:abun_errors}

Derived abundances are affected by random (internal) and systematic (external) errors. 

Sources of internal errors include uncertainties in atomic and stellar parameters, measured equivalent widths, 
and the veiling determination. 

Uncertainties in atomic parameters, such as the transition probability ($\log gf$), should cancel out, since our analysis is carried out differentially 
with respect to the Sun. 

Errors due to uncertainties in stellar parameters ($T_{\rm eff}$, $\xi$, $\log g$) were estimated first by 
assessing errors in stellar parameters 
themselves and then by varying each parameter separately, while keeping the other two unchanged. We found that variations in $T_{\rm eff}$ larger 
than 60~K would introduce spurious trends in $\log n{\rm (Fe)}$ versus the excitation potential ($\chi$), while variations in $\xi$ larger 
than 0.2 km s$^{-1}$ would result in significant trends of $\log n{\rm (Fe)}$ versus EW, and variations in $\log g$ larger than 0.2 dex 
would lead to differences between $\log n{\rm (\ion{Fe}{i})}$ and $\log n{\rm (\ion{Fe}{ii})}$ larger than 0.05 dex. The above values were thus 
assumed as uncertainties in stellar parameters. Errors in abundances (both [Fe/H] and [X/H]) due to uncertainties in stellar 
parameters are summarized in Table~\ref{tab:errors} for one of the coolest and the warmest stars in our ONC/OB1b samples. 

As for the errors due to uncertainties in EWs, our spectra are characterized by different $S/N$ ratios and it is not possible {\it a priori} 
to estimate a typical error in EW. However, random errors in EW are well represented by the standard deviation around the mean abundance 
determined from all the lines. These errors are listed in Table~\ref{tab:param_chem}, where uncertainties in [X/Fe] were obtained by 
quadratically adding the [Fe/H] error and the [X/H] error. When only one line was measured, the error in [X/H] is the standard deviation of 
three independent EW measurements. The number of lines employed for the abundance analysis is listed in Table~\ref{tab:param_chem} in 
parentheses, with the exception of those elements (sodium, aluminium, and ionized titanium) where only one or two lines were used.

Finally, as described in Sect.~\ref{sec:veiling}, the veiling of all the sample stars is negligible, with the 
only exception of JW373. For this star, we estimate that the veiling contribution to the abundance error is 
on the order of 0.05-0.11 dex, depending on the element. 

The greatest contribution to the external or systematic errors originates in the abundance scale, which is 
mainly affected by the choice of the model atmospheres. This error source is discussed in Appendix 
\ref{appendix:b}. Here, we emphasize that we used for both the Orion subgroups the same procedure, instrument set-up, and prescriptions to derive the abundances. As a consequence, differences in 
abundance between ONC and OB1b should not be influenced by systematic errors linked to the abundance scale. 

\begin{table*}  
\caption{Internal errors in abundance determination due to uncertainties in stellar parameters for one of the coolest 
star (namely, CVSO118) and for the warmest star (namely, P1455) in our samples. Numbers refer to the differences 
between the abundances obtained with and without the uncertainties in stellar parameters.}
\label{tab:errors}
%\tiny
\begin{center}
\begin{tabular}{lccc}
\hline
\hline
CVSO118 & $T_{\rm eff}=4300$ K & $\log g=4.3$ & $\xi=1.5$ km/s\\
\hline
$\Delta$   & $\Delta T_{\rm eff}=-/+60$ K & $\Delta \log g=-/+0.2$ & $\Delta \xi=-/+0.2$ km/s\\
\hline
$[$\ion{Fe}{i}/H$]$  & 0.02/$-$0.02 & $-$0.02/0.02 & 0.05/$-$0.05  \\
$[$Na/Fe$]$ & $-$0.06/0.07 & 0.07/$-$0.07 & $-$0.03/0.03     \\
$[$Al/Fe$]$ & $-$0.03/0.04 & 0.03/$-$0.02 & $-$0.03/0.03  \\
$[$Si/Fe$]$ & 0.02/$-$0.06 & $-$0.06/0.01 & $-$0.07/0.02  \\
$[$Ca/Fe$]$ & $-$0.07/0.07 & 0.07/$-$0.08 & $-$0.01/0.00     \\
$[$\ion{Ti}{i}/Fe$]$ & $-$0.07/0.08 & 0.03/$-$0.03 & 0.05/$-$0.04  \\
$[$\ion{Ti}{ii}/Fe$]$& 0.03/$-$0.03 & $-$0.08/0.07 & $-$0.02/0.01  \\
$[$Ni/Fe$]$ & 0.01/$-$0.01 & $-$0.03/0.03 & $-$0.02/0.02  \\				       
\hline	\\	
P1455 & $T_{\rm eff}=5950$ K & $\log g=4.1$ & $\xi=1.8$ km/s\\
\hline
        & $\Delta T_{\rm eff}=-/+60$ K & $\Delta \log g=-/+0.2$ & $\Delta \xi=-/+0.2$ km/s\\
\hline
$[$\ion{Fe}{i}/H$]$ & $-$0.05/0.03 & 0.00/$-$0.02 & 0.05/$-$0.06\\
$[$\ion{Fe}{ii}/H$]$& 0.05/0.00 & $-$0.06/0.11 & 0.07/$-$0.02\\
$[$Na/Fe$]$ & 0.01/0.00 & 0.05/$-$0.04 & $-$0.02/0.01\\
$[$Al/Fe$]$ & 0.02/0.00 & 0.01/0.01 & $-$0.04/0.05\\
$[$Si/Fe$]$ & 0.04/$-$0.02 & 0.00/0.03 & $-$0.04/0.05\\
$[$Ca/Fe$]$ & 0.01/0.01 & 0.02/0.00 & $-$0.01/0.02\\
$[$\ion{Ti}{i}/Fe$]$ & $-$0.01/0.03 & 0.01/$-$0.01 & $-$0.03/0.05\\
$[$\ion{Ti}{ii}/Fe$]$& 0.05/$-$0.04 & $-$0.09/0.10 & $-$0.02/0.03\\
$[$Ni/Fe$]$ & 0.01/0.01 & 0.00/0.02 & $-$0.02/0.03\\
\hline	\\	
\end{tabular}
\end{center}
\end{table*}

\section{Results}
\label{sec:results}

\subsection{Metallicity}
\begin{figure}	%[b!]
\begin{center}
 \begin{tabular}{c}
  \resizebox{\hsize}{!}{\includegraphics{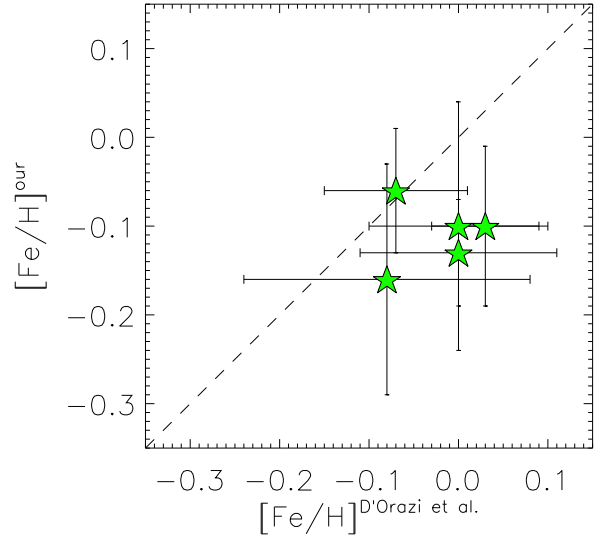} } \\
 \end{tabular}
       \caption{Our iron abundance measurements of the \cite{dorazietal09} sample versus their outputs.}
       \label{fig:re-analysis}
 \end{center}
\end{figure}

\begin{figure}[b!]
\begin{center}
 \begin{tabular}{c}
  \resizebox{\hsize}{!}{\includegraphics{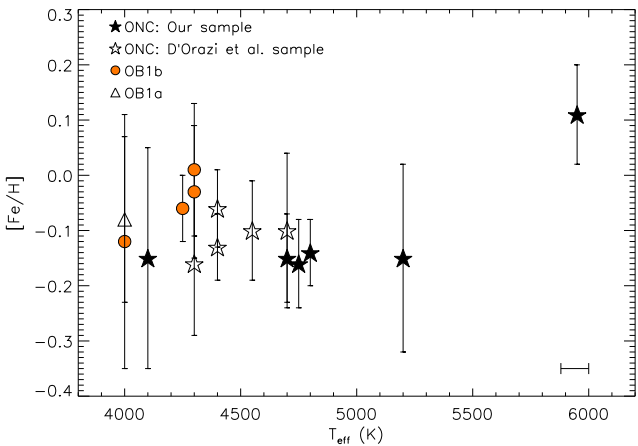}}\\                 
 \end{tabular}
       \caption{[Fe/H] versus $T_{\rm eff}^{\rm S}$ for the OB1b targets and the ONC stars observed by us and 
       by \cite{dorazietal09} and reanalyzed by ourselves. The position of CVSO56 (OB1a) is also shown. The 
       horizontal bar represents the typical uncertainty in $T_{\rm eff}^{\rm S}$. }
       \label{fig:feh_teff}
 \end{center}
\end{figure}

\begin{figure*}	%[b!]
\begin{center}
 \begin{tabular}{c}
  \resizebox{\hsize}{!}{\includegraphics{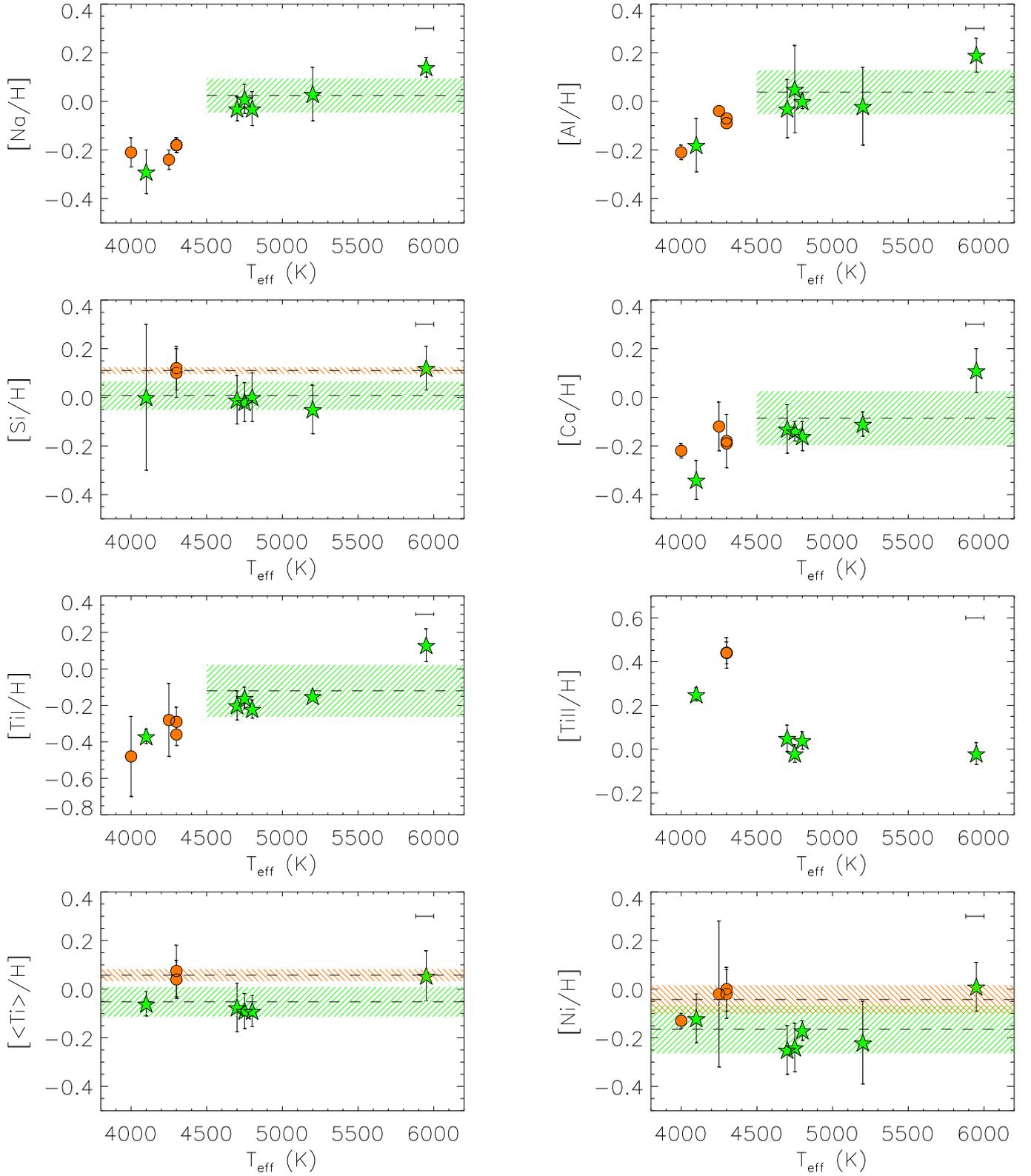}}\\   
\end{tabular}
       \caption{[X/H] versus $T_{\rm eff}^{\rm S}$. Stars and circles represent ONC and OB1b targets, respectively. Mean ONC 
       and OB1b values and $\pm 1\sigma$ bars are shown as dashed areas, $45\degr$ and $215\degr$ oriented, respectively. 
       For the Na, Al, Ca, and Ti abundances, we show the mean values obtained from all the ONC targets with $T_{\rm eff}^{\rm S}>4500$ K, 
       while for the Si, Ni, and mean Ti abundances the average was computed considering all the ONC/OB1b stars. The horizontal error bar in 
       all plots represents the typical uncertainty in $T_{\rm eff}^{\rm S}$.}
       \label{fig:xh_teff}
 \end{center}
\end{figure*}

\begin{figure*}	%[b!]
\begin{center}
 \begin{tabular}{c}
  \resizebox{\hsize}{!}{\includegraphics{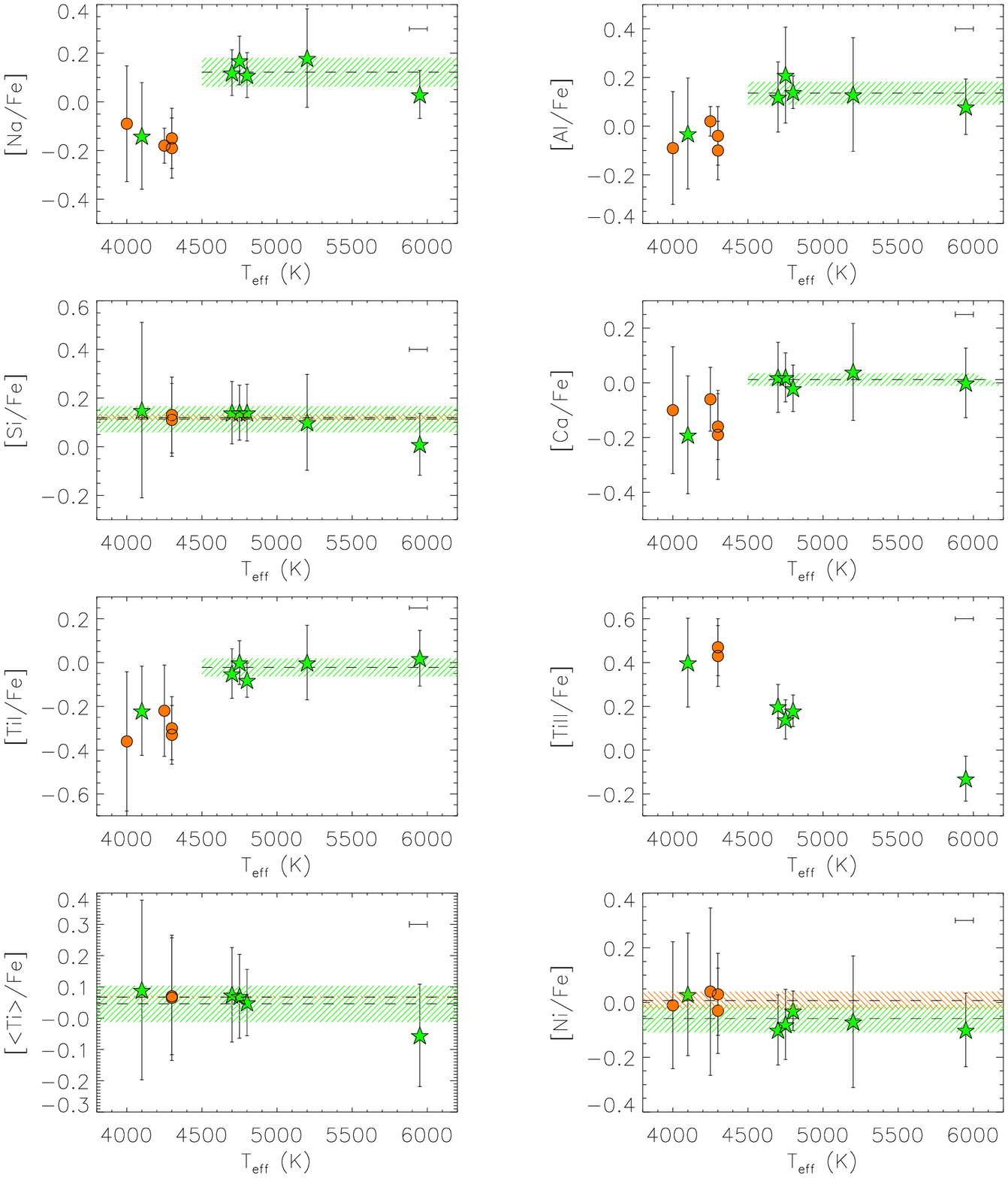}}\\   
\end{tabular}
       \caption{[X/Fe] versus $T_{\rm eff}^{\rm S}$. Symbols as in Fig.~\ref{fig:xh_teff}.}
       \label{fig:xfe_teff}
 \end{center}
\end{figure*}

Our final abundances are listed in Table~\ref{tab:param_chem}. The comparison between the new and the \cite{dorazietal09} iron 
abundances is given in Table~\ref{tab:re-analysis} and shown in Fig.~\ref{fig:re-analysis}. We comment on any 
differences in Sect.~\ref{sec:previous_results}. In Figs.~\ref{fig:feh_teff}, \ref{fig:xh_teff}, and 
\ref{fig:xfe_teff}, we present [Fe/H], [X/H], and [X/Fe] as a function of $T_{\rm eff}$ for the ONC and 
OB1b samples. Figure~\ref{fig:feh_teff} includes both our targets and those from \cite{dorazietal09}. 

Figure~\ref{fig:feh_teff} shows that, with the only exception of P1455 in the ONC, we do not find a major star-to-star difference in [Fe/H], 
with a remarkable agreement between the iron abundance of the present sample and that of D'Orazi et al. Excluding 
the D'Orazi et al. sample, the mean ONC [Fe/H] is $-0.15\pm0.01$ (without P1455) and $-0.11\pm0.11$ (with P1455). Including those stars, 
we find similar values, namely $-0.13\pm0.03$ and $-0.11\pm0.08$, respectively. As for OB1b, the star-to-star difference is minimal, considering 
the rather large uncertainty that affects the measurement of the coolest star. The mean for OB1b is [Fe/H]$=-0.05\pm0.05$, 
i.e. almost 0.1 dex above the ONC, although marginally consistent with it. 

Moreover, Fig.~\ref{fig:feh_teff} does not reveal any trend between [Fe/H] and effective temperature, with the exception again of P1455, the 
warmest star of the sample. Therefore, we believe that the difference between the [Fe/H] values of the ONC and OB1b is not due to 
systematic effects related to the effective temperature.

Finally, we mention that for the likely OB1a/25~Ori member (CVSO56) we find [Fe/H]=$-0.08\pm0.15$, a value very close to that of the ONC 
and OB1b (see Fig.~\ref{fig:feh_teff}).

\subsubsection{[Fe/H] difference between ONC and OB1b?}
As mentioned in the Introduction, \cite{dorazietal09} found that the only star of OB1b (HD294297) was $\sim 0.1$~dex more metal-poor than the ONC. 
\cite{caballero2010} demonstrated that it is a non-member of the association. Our new results show that the average [Fe/H] 
of OB1b is slightly higher than that of ONC. The question then arises of whether the ONC is intrinsically more 
metal-poor than OB1b or there are systematic effects in the analysis.

\begin{figure}[b!]
\begin{center}
 \begin{tabular}{c}
  \resizebox{\hsize}{!}{\includegraphics{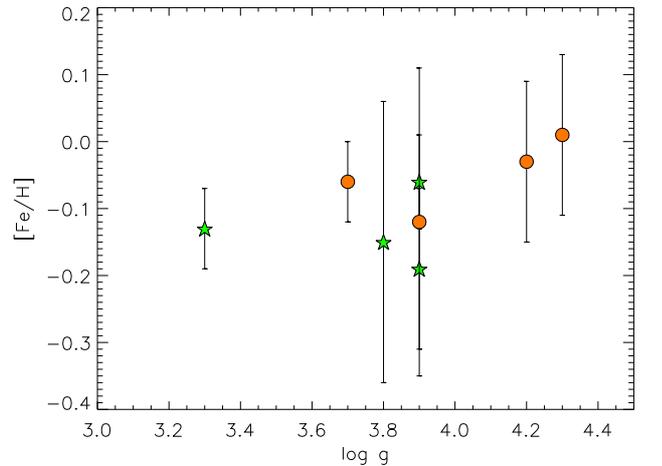}}\\                 
 \end{tabular}
       \caption{[Fe/H] versus $\log g$ for the coolest stars ($T_{\rm eff} < 4500$ K) of our and D'Orazi et al. re-analyzed samples 
       (stars: ONC; circles: OB1b).}
       \label{fig:feh_logg}
 \end{center}
\end{figure}

We note that NLTE effects have indeed been found to be important for cool stars with relatively high gravity, 
leading to an overestimate of the [Fe/H] (\citealt{takeda2008,schuleretal2010}). In Fig.~\ref{fig:feh_logg}, we 
show [Fe/H] versus $\log g$ for our ONC/OB1b stars cooler than 4500 K; the figure shows a moderate increase 
in the abundance with $\log g > 4.0$. Although we cannot quantitatively estimate the amount of NLTE effects, 
we suggest that they might account for the difference between the ONC and OB1b metallicities. 

\subsection{Other elements}
In Figs.~\ref{fig:xh_teff} and \ref{fig:xfe_teff}, we plot [X/H] and [X/Fe] as a function of $T_{\rm eff}$ for 
the different elements derived in this study. These figures show that no trends are present for silicon and nickel, 
with the only exception of P1455, which is a `rich' star in terms of all elements, included iron (see 
Sect.~\ref{sec:p1455}). Its values of [X/Fe] are instead close to the solar ones (Fig.~\ref{fig:xfe_teff}). 
In contrast, Na, Al, Ca, and \ion{Ti}{i} are lower for stars cooler than $\sim$4500 K both when considering 
[X/H] and when considering [X/Fe]. 

Similar results have been found by several authors (e.g., \citealt{Schuleretal2003, Schuleretal2006, Yongetal2004, Beirao05, Gillietal2006, dorazirandich09}) 
and ascribed to NLTE effects. In particular, \cite{dorazirandich09} in their analysis of the young clusters IC~2602 and IC~2391 
show that, while the [\ion{Ti}{i}/Fe] ratio decreases with decreasing temperature, [\ion{Ti}{ii}/Fe] increases, suggesting that 
over-ionization is at work. Owing to their young age, cluster stars are characterized by enhanced levels of chromospheric activity and are more affected 
by NLTE over-ionization. As suggested by the aforementioned authors, deviations from LTE are possibly responsible for the effects 
seen in Fig.~\ref{fig:xfe_teff} for Na, Al, Ca, and Ti elements. On the other hand, Ni and Si lines have higher ionization 
potentials (7.63 and 8.15 eV, respectively) than Na, Al, Ca, and Ti ($\sim 5.14-6.82$ eV) and are thus less affected by over-ionization.

Regardless of the physical reasons for the observed behavior, we computed the mean [X/Fe] ratios for sodium, 
aluminum, and calcium considering only the ONC/OB1b stars with temperatures higher than 4500 K (Table~\ref{tab:param_chem}). 
In the case of titanium, where some ionized lines were measured, we list the average coming from \ion{Ti}{i} and \ion{Ti}{ii}. There is no 
significant difference in the average values between the two subgroups and the dispersion observed in the elemental abundances of each subgroup 
is smaller than the observational uncertainties. The average [X/Fe] values are close to solar, with Na, Al, and Si being slightly overabundant.

\subsection{Comparison with results of other authors}
\label{sec:previous_results}
Some of our sample stars have been observed by other authors. In particular, one star, JW157=KM~Ori, is in common with the \cite{padgett1996} 
sample. She finds for this star [Fe/H]=$+0.14\pm0.18$, which disagrees significantly with our iron abundance ($\Delta$[Fe/H]=0.29 dex). Possible 
explanations include: 
\begin{itemize}
\item[$i)$] The Padgett abundance analysis was based on fewer iron lines (17) than our own (40 lines; see 
Table~\ref{tab:param_chem}); 
\item[$ii)$] whereas we excluded very strong lines with $EW > 150$ m\AA~(see Sect.~\ref{sec:lin_sol_ew}), 6 among 
17 lines in the Padgett list have $EW > 150$ m\AA~(see her Table 9); 
\item[$iii)$] seven lines of the Padgett list are in common with us, but for four of those we find different EWs 
of about $\pm$15-20 m\AA. We carefully re-measured these lines confirming our original values. We conclude that 
our determination is likely to be more reliable. 
\end{itemize}

Both the mean [Fe/H] value for the ONC and [X/Fe] ratios found by ourselves are in good agreement with the 
results of \cite{santosetal2008}, who derived a mean metallicity of [Fe/H]=$-0.13\pm0.06$ and Ni and Si 
abundances [Ni/Fe]=$-0.06 \pm 0.07$ and [Si/Fe]$=0.00\pm 0.09$, respectively. In particular, two of three 
stars of their sample (JW589 and JW157) are in common with ours. There is a reasonable agreement between the 
abundances for these two stars, and any differences can be accounted for by the different line lists and 
$\sigma$-clipping criteria used. 

As shown in Sect.~\ref{sec:parameters}, the present results yield a lower [Fe/H] than \cite{dorazietal09}. 
The main reasons for this discrepancy are: 
\begin{itemize}
\item[$i)$] our new analysis is based on more suitable spectra and line list of \ion{Fe}{i}, which allow us to 
more tighly constrain the abundance versus $\chi$ and versus $EW$ trends, hence the $T_{\rm eff}$ and $\xi$ values; 
\item[$ii)$] our analysis of stars with $T_{\rm eff} \ltsim 4400$ K is based on GAIA atmospheric models that are 
more appropriate for cool stars (as discussed in Appendix~\ref{appendix:b}); 
\item[$iii)$] critical and careful reanalysis of the CD\#4 spectra acquired by D'Orazi et al. allowed us to 
derive the microturbulence value of the star \# 673, which had been not inferred by these authors. 
\end{itemize}

\section{Discussion}
\label{sec:discussion}

\subsection{Triggered star formation and chemical self-enrichment}

Orion is regarded as a proto-type of triggered star formation, where star formation has proceeded sequentially 
(\citealt{preibzinne2006}). As mentioned in the Introduction, this scenario predicts a peculiar chemical enrichment 
due to contamination of material ejected from type-II supernovae (SNIIe) originating from a first generation 
of massive stars, since these are expected to contain the nucleosynthetic products of the stellar interior.

In support of this view, \cite{cunhalamb92,cunhalamb94} found that stars in the young subgroup 1d and some of the 
slightly older subgroup 1c have an abundance up to about 40\% higher than the rest of their sample. They suggested 
that the enrichment resulted from the mixing of SNIIe ejecta from the 1c subgroup to the center of the Trapezium 
cluster. \cite{simondiaz2010} derived homogeneous values of the oxygen and silicon abundances in stars of the 
four subgroups (OB1a,b,c,d), which had a dispersion ($\sim$0.04 dex) smaller than the intrinsic uncertainties 
($\sim \pm$0.10 dex). 

Our results indicate that low-mass stars yield the same abundance distribution as high-mass stars (see \citealt{simondiaz2010}). 
In particular, Si is the only $\alpha$-element that does not exhibit strong evidence of being affected by NLTE 
on the basis of our data (see Fig.~\ref{fig:xfe_teff}) and for which we obtained abundances for both ONC-OB1d 
and OB1b. We find group-to-group dispersions of $\sim$0.08 and $\sim$0.01 dex in [Si/H] and [Si/Fe], respectively, 
which are smaller than our internal errors (of around $\pm$0.11-0.31 and $\pm$0.13-0.36 dex, respectively). 
The other elements for which we measured the abundance is titanium. We find for this element a difference of 
$0.08$ dex for $<$[Ti/H]$>$ and $0.01$ dex for $<$[Ti/Fe]$>$; this is smaller than our internal errors (around 
$\pm$0.10-0.15 and $\pm$0.12-0.31 dex, respectively). 
 
We conclude that the ONC and OB1b are characterized by homogeneous silicon and titanium abundances. This means that even if SNII explosions 
occurred in OB1b, at the OB1b-ONC distance their ejecta did not have the conditions to chemically enrich the ONC stars, dispersing 
the element over a large volume.

As for the difference in [Fe/H] between the ONC and OB1b (should it be real), we note that an inhomogeneity in 
metallicity (at the level of $\sim 0.05$ dex) within a given star-forming complex is expected in models of 
hierarchical star formation (\citealt{elmegreen1998}). Owing their chaotic and large-scale formation process on 
a 1 kpc scale, the gas in a giant molecular cloud will have a range of metallicities reflecting the background 
Galactic gradient. We find this unlikely because the separation between ONC and OB1b ($< 50$ pc) is much smaller 
than the scale on which the Galactic gradient operates. 

\subsection{The metallicity of SFRs: comparison with young open clusters}
Our present analysis reinforces the conclusion of \cite{santosetal2008} that none of the SFRs with available 
metallicity is more metal-rich than the Sun; the majority of them are indeed slightly more metal-poor. 
\cite{santosetal2008} suggest that, if the lower-than-solar metallicities of SFRs were confirmed, this would 
imply that either the Sun was formed in an inner region of the Milky Way disk, or that the nearby interstellar 
medium experienced a recent infall of metal-poor gas.

\begin{figure}[b!]
\begin{center}
 \begin{tabular}{c}
  \resizebox{\hsize}{!}{\includegraphics[angle=-90]{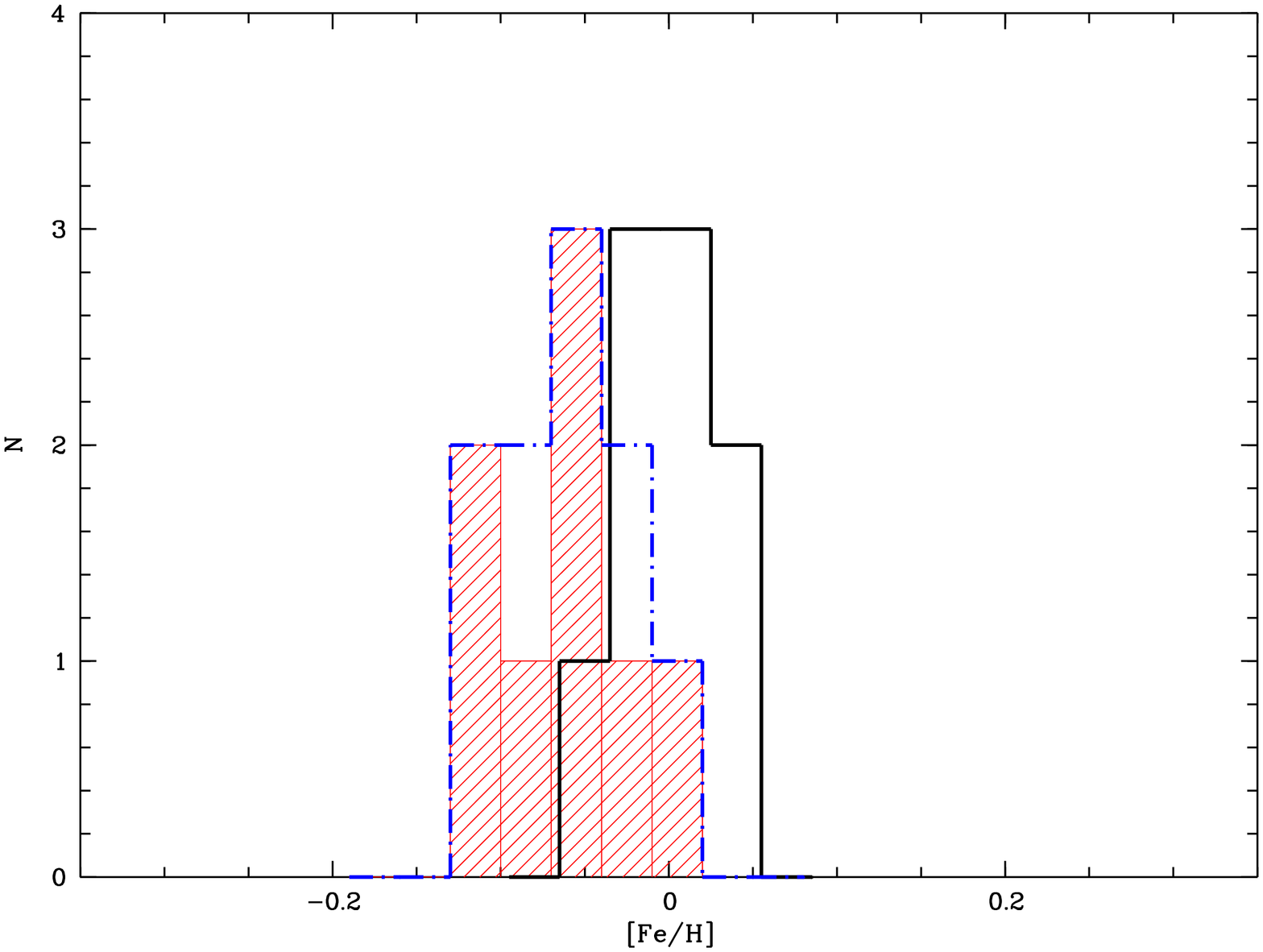}}\\
 \end{tabular}
       \caption{[Fe/H] distribution for: SFRs within 500~pc from the Sun (dashed histogram), young nearby loose 
       associations (dot-dashed line), and open clusters younger than 150~Myr and within 500~pc from the Sun 
       (solid line). Sources of [Fe/H] values for the young open clusters are the following: $\alpha$~Persei 
       (\citealt{boesgaardfriel1990}), NGC~2516 (\citealt{terndrupetal2002}), NGC2451A/B (\citealt{hunschetal2004}), 
       Blanco~1 (\citealt{fordetal2005}), IC~4665 (\citealt{shenetal2005}), IC~2602 and IC~2391 
       (\citealt{dorazirandich09}), and Pleiades (\citealt{soderblometal2009}).
}
\label{fig:histo}
 \end{center}
\end{figure}

\cite{dorazirandich09} demonstrated that the abundance pattern of young open clusters in the solar neighborhood 
is identical to the solar distribution, concluding that the Sun was most likely born at the present location. 
In Fig.~\ref{fig:histo}, we compare the distribution of [Fe/H] of the SFRs within 500~pc of the Sun with that 
of $i)$ open clusters younger than $\sim 150$~Myr within the same distance from the Sun, and $ii)$ young 
nearby loose associations. In addition to the [Fe/H] values for the ONC and OB1b derived here, metallicity 
determinations for other SFRs were retrieved from Santos et al. (2008 - Chamaeleon, $\rho$~Ophiucus, Corona 
Australis, Lupus), Gonz\`alez-Hern\`andez et al. (2008 - $\sigma$~Orionis), and D'Orazi et al. (in 
preparation - Taurus). [Fe/H] values for the young associations were taken from Viana Almeida et al. 
(2009 - their `uncorrected' values were considered).

The figure shows that the three distributions, in particular that of open clusters, are characterized by a small dispersion; 
however, the cluster distribution is shifted towards somewhat higher metallicities than the SFRs and young associations. 
We obtain average values of [Fe/H]$=-0.06\pm 0.04, -0.06\pm 0.04$, and $-0.01\pm 0.03$ for the SFRs, associations, and open clusters, respectively. 
In addition, none of the open clusters is as metal-poor as the most metal-poor SFR (the ONC) and only three out of nine SFRs fall within the 
open cluster distribution. In other words, not only the SFRs are more metal-poor than the Sun, but they are on average more metal-poor 
than young open clusters, which should be representative of the metallicity in the solar neighborhood. As for the loose 
associations, their distribution is in closer agreement with that of the SFRs than with the clusters; we note, however, that the 
`corrected' values of \cite{vianaetal2009} would yield a higher metallicity. Higher values of [Fe/H] were also derived by 
\cite{rojasetal2008} for a sample of Tucana-Horologium members. 

Focusing on the comparison between clusters and SFRs, the offset between their mean [Fe/H] values is small and still 
based on relatively small number statistics. One might have missed the more metal-rich SFRs or, viceversa, the more 
metal-poor open clusters in the solar vicinity. In addition, whereas an agreement on the metallicity of the ONC 
seems now to have been reached, discrepancies still exist for other regions; for example, \cite{santosetal2008} 
find [Fe/H]$=-0.08\pm 0.12$ for Ophiucus, compared to the value [Fe/H]$=0.08\pm 0.07$ derived by 
\cite{padgett1996}. This indicates that additional homogeneous studies must be performed, before the conclusion 
that SFRs are slightly more metal-poor than the Sun and the open clusters can be definitively drawn.

With this caveat in mind, we would like to point out that, given the young age of the clusters, stellar migration 
is unlikely to be the reason for the difference in the [Fe/H] distributions of the open clusters and SFRs. The 
difference likely reflects a difference in the interstellar gas from which members of SFRs and young clusters 
formed. This in turn must be a relic of the process of star formation in the solar neighborhood, rather than 
an effect of chemical evolution, given the short timescales involved, and that in any case chemical evolution 
would lead to the younger regions (i.e., the SFRs) being more metal-rich than the older clusters.

\subsection{The case of P1455: a metal-rich star?}
\label{sec:p1455}
P1455 is significantly more metal-rich than the ONC stars. For this target, we derived an iron abundance of 
$0.11\pm0.09$ in very good agreement with the \cite{cunhaetal95} value of [Fe/H]=$0.08\pm0.15$, which confirms 
that this star is more metal-rich than the other ONC targets at the 2-$\sigma$ level. Although this is still 
marginally consistent with what is expected from statistical fluctuations, some discussion of this object 
would be merited. 

At present, we do not have reasons to consider this star as a non-member of the ONC. Its radial velocity and 
proper motion are consistent with membership. We also note that we obtained an independent RV estimate 
using spectra acquired with {\sc harps} (High Accuracy Radial velocity Planet Searcher; \citealt{mayoretal2003}), 
yielding a mean $V_{\rm rad}$ of 21.893$\pm$0.014 km s$^{-1}$.

As noted in Sect.~\ref{sec:sample}, this star is farther away from the main cluster than the other targets. 
Its higher metallicity may imply that the ONC region is inhomogeneous on scales of 1-2 pc and that one could 
search for other metal-rich stars close to P1455.

We recall that whether super-solar metallicity stars exists in SFRs has been disputed for a few years 
(\citealt{santosetal2008}, and references therein), and is relevant because circumstellar disks around young 
stars are the birthplace of planets. Thus, `P1455-like cases' in SFRs may well be good targets for exoplanet 
searches.

\section{Conclusions}
\label{sec:conclusion}
We have presented new measurements of the abundances of iron-peak elements and $\alpha$-elements in two subgroups 
of the Orion complex, the Orion Nebula Cluster (ONC), and the OB1b sub-association, derived from {\sc flames/uves} 
high-resolution spectroscopy. Our main results can be summarized as follows:
\begin{itemize}
\item The ONC and OB1b have mean iron abundances of $-0.11\pm0.08$ and $-0.05\pm0.05$, respectively. A likely 
member of OB1a has an abundance of [Fe/H]=$-0.08\pm0.15$. We can exclude the metallicity of Orion being above 
the solar value.
\item The ONC and OB1b are characterized by a small scatter in iron abundances, with the only exception of 
P1455 in the ONC, which we confirm to be metal-rich, as found in previous studies (\citealt{cunhaetal98, cayrel2001}). 
\item In the temperature range where NLTE effects are less evident (i.e. $\gtsim 4500$ K), there is no 
presence in the ONC sample of star-to-star inhomogeneity in the abundances of the elements strongly affected 
by these effects (namely Na, Al, Ca, \ion{Ti}{i}). Owing to the lower temperatures of the OB1b sample, we are 
unable to draw any conclusions about these elements. For elements not strongly affected by NTLE effects 
(namely, Si and Ni) both the ONC and OB1b do not show 
any star-to-star abundance inhomogeneity.
\item The two sub-associations analyzed here have similar solar abundances of the $\alpha$-elements silicon 
and titanium (the latter obtained by averaging the abundance of \ion{Ti}{i} and \ion{Ti}{ii}). Similar nickel 
abundances were found for the two Orion subgroups. No evidence of self-enrichment from OB1b to the ONC is found. 
\item Star-forming regions and open clusters younger than 150 Myr and within 500~pc of the Sun were found 
to have a small offset between their mean iron abundance, with the former being more metal-poor than the 
latter. Owing to the young age of both sets of stars, this offset probably reflects a difference in the 
properties of the interstellar gas from which members of SFRs and young clusters formed. More homogeneous 
studies are required to draw definitive conclusions.
\end{itemize}

\begin{acknowledgements}
The authors are very grateful to the referee for a careful reading of the paper and constructive suggestions.
Jacob Bean provided some IDL codes, for which we are grateful. We thank Daniele Galli, Laura Magrini, and Fabrizio 
Massi for very useful discussions. This research has made use of the SIMBAD database, operated at CDS (Strasbourg, France), 
and of the WEBDA database, operated at the Institute for Astronomy of the University of Vienna.
\end{acknowledgements}

\bibliographystyle{aa}

%\newpage

\appendix{
\section{Line list}
\label{appendix:a}

\begin{longtable}{rrrr}  
\caption{Wavelength, elements, excitation potential, and oscillator strength of all the elements are listed.\label{tab:line_list}}\\
%\tiny
\hline		
\hline
\multicolumn{1}{r}{$\lambda$}
&\multicolumn{1}{r}{Element}
&\multicolumn{1}{r}{$\chi$}
&\multicolumn{1}{r}{$\log gf$}\\
\multicolumn{1}{r}{(\AA)}
&\multicolumn{1}{r}{}
&\multicolumn{1}{r}{(eV)}
&\multicolumn{1}{r}{}\\
\hline
\endfirsthead 
\multicolumn{4}{c} 
{{Table \ref{tab:line_list}: } 
(\textit{continued})}\\[5pt] 
\hline 
\multicolumn{1}{r}{$\lambda$}
&\multicolumn{1}{r}{Element}
&\multicolumn{1}{r}{$\chi$}
&\multicolumn{1}{r}{$\log gf$}\\
\multicolumn{1}{r}{(\AA)}
&\multicolumn{1}{r}{}
&\multicolumn{1}{r}{(eV)}
&\multicolumn{1}{r}{}\\
\hline
\endhead 
\hline
\multicolumn{4}{r}{\small\itshape continued on the next page}\\
\endfoot
%\hline 
\endlastfoot
%5682.633 &  \ion{Na}{i} &2.102 & $-$0.700 \\
5688.205 &  \ion{Na}{i} &2.104 & $-$0.452 \\
6154.226 &  \ion{Na}{i} &2.102 & $-$1.610 \\
6160.747 &  \ion{Na}{i} &2.104 & $-$1.310 \\
6696.023 &  \ion{Al}{i} &3.143 & $-$1.499 \\
6698.673 &  \ion{Al}{i} &3.143 & $-$1.950 \\
%5701.104 &  \ion{Si}{i} &4.930 & $-$2.050 \\ 
5948.541 &  \ion{Si}{i} &5.082 & $-$1.230 \\ 
6091.919 &  \ion{Si}{i} &5.871 & $-$1.400  \\ 
6125.021 &  \ion{Si}{i} &5.614 & $-$1.570 \\ 
6142.483 &  \ion{Si}{i} &5.619 & $-$1.480 \\ 
6145.016 &  \ion{Si}{i} &5.616 & $-$1.440 \\ 
6414.980 &  \ion{Si}{i} &5.871 & $-$1.100  \\ 
%6518.733 &  \ion{Si}{i} &5.954 & $-$1.500  \\ 
6555.463 &  \ion{Si}{i} &5.984 & $-$1.000  \\ 
5512.980 &  \ion{Ca}{i} &2.933 & $-$0.480  \\ 
5581.965 &  \ion{Ca}{i} &2.523 & $-$0.671 \\ 
%5601.277 &  \ion{Ca}{i} &2.526 & $-$0.523 \\ 
5867.562 &  \ion{Ca}{i} &2.933 & $-$1.610  \\ 
%6163.754 &  \ion{Ca}{i} &2.521 & $-$1.290  \\ 
6166.439 &  \ion{Ca}{i} &2.521 & $-$1.156 \\ 
%6169.042 &  \ion{Ca}{i} &2.523 & $-$0.804 \\ 
%6169.563 &  \ion{Ca}{i} &2.526 & $-$0.527 \\
6455.598 &  \ion{Ca}{i} &2.523 & $-$1.424 \\
6499.650 &  \ion{Ca}{i} &2.523 & $-$0.818 \\
%6572.795 &  \ion{Ca}{i} &1.886 & $-$4.320  \\
%4805.415 &  \ion{Ti}{i} &2.345 &    0.150 \\ 
4820.411 &  \ion{Ti}{i} &1.502 & $-$0.441 \\ 
%4885.079 &  \ion{Ti}{i} &1.887 &    0.358  \\
%4913.614 &  \ion{Ti}{i} &1.873 &    0.160   \\
%5016.161 &  \ion{Ti}{i} &0.848 & $-$0.574 \\ 
5219.702 &  \ion{Ti}{i} &0.021 & $-$2.292 \\ 
5866.451 &  \ion{Ti}{i} &1.067 & $-$0.840  \\ 
5953.160 &  \ion{Ti}{i} &1.887 & $-$0.329 \\ 
5965.828 &  \ion{Ti}{i} &1.879 & $-$0.409 \\ 
6126.224 &  \ion{Ti}{i} &1.067 & $-$1.424 \\ 
6258.102 &  \ion{Ti}{i} &1.443 & $-$0.431 \\ 
6261.098 &  \ion{Ti}{i} &1.430 & $-$0.479 \\
6743.127 &  \ion{Ti}{i} &0.900 & $-$1.630  \\ 
6491.560 &  \ion{Ti}{ii}&2.061 & $-$1.793 \\   
6559.590 &  \ion{Ti}{ii}&2.048 & $-$2.019 \\   
6680.133 &  \ion{Ti}{ii}&3.095 & $-$1.855 \\   
4835.868 &  \ion{Fe}{i} &4.103 & $-$1.500  \\ 
4875.878 &  \ion{Fe}{i} &3.332 & $-$2.020  \\
4907.732 &  \ion{Fe}{i} &3.430 & $-$1.840  \\
5044.211 &  \ion{Fe}{i} &2.851 & $-$2.059 \\
5141.739 &  \ion{Fe}{i} &2.424 & $-$2.190  \\
%5162.273 &  \ion{Fe}{i} &4.178 &    0.020  \\
%5217.389 &  \ion{Fe}{i} &3.211 & $-$1.070  \\
%5228.377 &  \ion{Fe}{i} &4.220 & $-$1.290  \\
5285.129 &  \ion{Fe}{i} &4.434 & $-$1.640 \\
5293.959 &  \ion{Fe}{i} &4.143 & $-$1.870  \\
5373.709 &  \ion{Fe}{i} &4.473 & $-$0.860  \\
5386.334 &  \ion{Fe}{i} &4.154 & $-$1.770  \\
5389.479 &  \ion{Fe}{i} &4.415 & $-$0.570  \\
5398.279 &  \ion{Fe}{i} &4.445 & $-$0.720  \\
5472.709 &  \ion{Fe}{i} &4.209 & $-$1.495 \\
5522.447 &  \ion{Fe}{i} &4.209 & $-$1.550  \\
5539.280 &  \ion{Fe}{i} &3.642 & $-$2.660  \\
5543.150 &  \ion{Fe}{i} &3.695 & $-$1.570  \\
5543.936 &  \ion{Fe}{i} &4.217 & $-$1.140  \\
%5546.992 &  \ion{Fe}{i} &4.217 & $-$1.910  \\
5576.089 &  \ion{Fe}{i} &3.430 & $-$0.894 \\
5584.765 &  \ion{Fe}{i} &3.573 & $-$2.320  \\
5636.696 &  \ion{Fe}{i} &3.640 & $-$2.610  \\
5638.262 &  \ion{Fe}{i} &4.220 & $-$0.870  \\
5641.434 &  \ion{Fe}{i} &4.256 & $-$1.063 \\
5691.497 &  \ion{Fe}{i} &4.301 & $-$1.520  \\
5701.545 &  \ion{Fe}{i} &2.559 & $-$2.216 \\
5856.088 &  \ion{Fe}{i} &4.294 & $-$1.570  \\
5859.578 &  \ion{Fe}{i} &4.549 & $-$0.620  \\
5862.353 &  \ion{Fe}{i} &4.549 & $-$0.365 \\
5916.247 &  \ion{Fe}{i} &2.453 & $-$2.994 \\
5930.180 &  \ion{Fe}{i} &4.652 & $-$0.251 \\
5934.655 &  \ion{Fe}{i} &3.928 & $-$1.170  \\
5956.694 &  \ion{Fe}{i} &0.859 & $-$4.605 \\
5976.775 &  \ion{Fe}{i} &3.943 & $-$1.290  \\
5984.814 &  \ion{Fe}{i} &4.733 & $-$0.280  \\
5987.066 &  \ion{Fe}{i} &4.795 & $-$0.556 \\
6003.012 &  \ion{Fe}{i} &3.881 & $-$1.120  \\
6024.058 &  \ion{Fe}{i} &4.548 & $-$0.052 \\
6056.005 &  \ion{Fe}{i} &4.733 & $-$0.460  \\
6078.491 &  \ion{Fe}{i} &4.795 & $-$0.370  \\
6157.728 &  \ion{Fe}{i} &4.076 & $-$1.260  \\
6187.990 &  \ion{Fe}{i} &3.943 & $-$1.720  \\
6200.313 &  \ion{Fe}{i} &2.608 & $-$2.450  \\
6315.811 &  \ion{Fe}{i} &4.076 & $-$1.710  \\
6322.685 &  \ion{Fe}{i} &2.588 & $-$2.446 \\
6330.850 &  \ion{Fe}{i} &4.733 & $-$1.158 \\
6336.824 &  \ion{Fe}{i} &3.686 & $-$0.856 \\
6344.149 &  \ion{Fe}{i} &2.433 & $-$2.923 \\
6469.193 &  \ion{Fe}{i} &4.835 & $-$0.770  \\
6495.742 &  \ion{Fe}{i} &4.835 & $-$0.940  \\
6498.939 &  \ion{Fe}{i} &0.958 & $-$4.699 \\
6574.228 &  \ion{Fe}{i} &0.990 & $-$5.023 \\
6609.110 &  \ion{Fe}{i} &2.559 & $-$2.692 \\
6627.545 &  \ion{Fe}{i} &4.548 & $-$1.500   \\
6703.567 &  \ion{Fe}{i} &2.758 & $-$3.100   \\
6713.745 &  \ion{Fe}{i} &4.790 & $-$1.410  \\
6725.364 &  \ion{Fe}{i} &4.100 & $-$2.210  \\
6726.673 &  \ion{Fe}{i} &4.610 & $-$1.050  \\
6733.151 &  \ion{Fe}{i} &4.638 & $-$1.580  \\
6750.164 &  \ion{Fe}{i} &2.420 & $-$2.580  \\
%6745.965 &  \ion{Fe}{i} &4.070 & $-$2.710  \\
6786.860 &  \ion{Fe}{i} &4.190 & $-$1.900  \\
6806.847 &  \ion{Fe}{i} &2.728 & $-$3.210  \\
6810.267 &  \ion{Fe}{i} &4.610 & $-$1.000   \\
5414.073 &  \ion{Fe}{ii}&3.221 & $-$3.750  \\
5425.257 &  \ion{Fe}{ii}&3.199 & $-$3.372 \\
5991.376 &  \ion{Fe}{ii}&3.153 & $-$3.560  \\
6084.111 &  \ion{Fe}{ii}&3.199 & $-$3.780  \\
6149.258 &  \ion{Fe}{ii}&3.889 & $-$2.800   \\
6247.557 &  \ion{Fe}{ii}&3.892 & $-$2.329 \\
6432.680 &  \ion{Fe}{ii}&2.891 & $-$3.685 \\
6456.383 &  \ion{Fe}{ii}&3.903 & $-$2.100   \\
6516.080 &  \ion{Fe}{ii}&2.891 & $-$3.450  \\
4806.984 &  \ion{Ni}{i} &3.679 & $-$0.640 \\
4852.547 &  \ion{Ni}{i} &3.542 & $-$1.070 \\
4904.407 &  \ion{Ni}{i} &3.542 & $-$0.170 \\
%4913.968 &  \ion{Ni}{i} &3.743 & $-$0.630  \\
5003.734 &  \ion{Ni}{i} &1.676 & $-$3.130\\
%5010.934 &  \ion{Ni}{i} &3.635 & $-$0.870 \\
5032.723 &  \ion{Ni}{i} &3.898 & $-$1.270 \\
5435.855 &  \ion{Ni}{i} &1.986 & $-$2.590 \\
5462.485 &  \ion{Ni}{i} &3.847 & $-$0.930 \\
5589.357 &  \ion{Ni}{i} &3.898 & $-$1.140 \\
5593.733 &  \ion{Ni}{i} &3.898 & $-$0.840 \\
%5625.312 &  \ion{Ni}{i} &4.089 & $-$0.700  \\
5641.881 &  \ion{Ni}{i} &4.105 & $-$1.080  \\
%5682.198 &  \ion{Ni}{i} &4.105 & $-$0.499 \\
5996.730 &  \ion{Ni}{i} &4.236 & $-$1.060 \\
6053.685 &  \ion{Ni}{i} &4.236 & $-$1.070 \\
6086.288 &  \ion{Ni}{i} &4.266 & $-$0.510 \\
6111.066 &  \ion{Ni}{i} &4.088 & $-$0.830 \\
6175.360 &  \ion{Ni}{i} &4.089 & $-$0.559\\
6186.709 &  \ion{Ni}{i} &4.105 & $-$0.960 \\
6191.171 &  \ion{Ni}{i} &1.676 & $-$2.353 \\
6204.604 &  \ion{Ni}{i} &4.088 & $-$1.140 \\
6223.981 &  \ion{Ni}{i} &4.105 & $-$0.970 \\
6327.604 &  \ion{Ni}{i} &1.676 & $-$3.150 \\
6378.247 &  \ion{Ni}{i} &4.154 & $-$0.830 \\
6384.668 &  \ion{Ni}{i} &4.154 & $-$1.130 \\ 
6586.308 &  \ion{Ni}{i} &1.951 & $-$2.810 \\ 
6598.611 &  \ion{Ni}{i} &4.236 & $-$0.980 \\ 
6635.137 &  \ion{Ni}{i} &4.419 & $-$0.830 \\ 
6767.784 &  \ion{Ni}{i} &1.830 & $-$2.060 \\  
6772.321 &  \ion{Ni}{i} &3.660 & $-$0.960 \\  
\hline		
%\end{tabular}
%\end{center}
\end{longtable}
%\normalsize
  
\section{Elemental abundance analysis of cool stars: dependence on model atmospheres}
\label{appendix:b}
Many possible fallacies can affect the process going from the spectroscopic stellar observations to the derivation 
of the chemical composition using atomic parameters and the derivation of stellar parameters. We focus here on the 
role of model atmospheres. In particular, we show how the use of different model atmospheres leads to different 
results in metallicity and other elemental abundances (i.e., sodium, aluminum, silicon, calcium, titanium, and 
nickel). 

\subsection{The test}
In the following, we present our starting points:
\begin{itemize}
\item We considered three young members of ONC and OB1b (namely, CVSO159, CVSO118, and KM~Ori) because they cover a wide range 
in effective temperature ($T_{\rm eff} \sim 4000-4700$ K) and surface gravity ($\log g\sim 3.0-4.5$), but our discussion can be extended 
to all late-G/early-M stars. We also considered, as a comparison, a solar 
spectrum acquired by \cite{randichetal2006} with {\sc flames/uves} at a similar resolution of the other spectra.
\item Abundance analysis was carried out following the steps given in Sect.~\ref{sec:abundance}.
\item We considered low-resolution (20 \AA) ATLAS\footnote{http://kurucz.harvard.edu/} (\citealt{Kuru93}) and 
high-resolution (2 \AA) GAIA\footnote{http://www.hs.uni-hamburg.de/EN/For/ThA/phoenix/} 
(\citealt{hauschildt1999,brotthauschildt10}) synthetic spectra to evaluate the continuum flux around lines and in 
photometric bands normally used for line EW measurements (see Sect.~\ref{sec:continuum}). ATLAS spectra cover the 
ultraviolet (1000 \AA) to infrared (10 $\mu$m) spectral range, while GAIA spectra cover the 300 \AA$ \ltsim \lambda \ltsim 100~\mu$m 
wavelength range.
\item \cite{Kuru93} and \cite{brotthauschildt10} grids of plane parallel model atmospheres were considered for the abundance 
measurements (see Sect.~\ref{sec:abundance_bis}). ATLAS includes atmosphere models with metallicities $-5.0 \le$[Fe/H]$\le +1.0$, 
gravity range $0.0\le \log g \le5.0$, and 3500$\le T_{\rm eff} \le$10\,000 K. GAIA model atmospheres span in 
$2000\le T_{\rm eff}\le 10000$ K, $0.0\le \log g \le 5.5$, and $-4.0 \le$[Fe/H]$\le +0.5$. Model atmospheres for specific 
stellar parameters of our interest were generated by interpolating in the original ATLAS and GAIA grids (see the procedure 
described by \citealt{beanetal2006}).
\end{itemize}

\subsection{Implications on continuum flux}
\label{sec:continuum}
One of the most important improvements made to the GAIA models was the inclusion of millions of molecular 
lines in the line list. This is of paramount importance when computing the band opacity, in addition to the 
line opacity. The effects of band opacities on the continuum flux are most pronounced in the optical domain 
that is largely used for abundance measurements (namely, 4000--8000 \AA). 

To help identify the range in effective temperature where the two grids of models can be used for abundance 
measurements, we calculated the GAIA average continuum fluxes in 20 \AA~windows centered on 
$\lambda$4900, 5600, 6300, and 7500 \AA, which are typical regions used for abundance determinations. These 
fluxes were evaluated for solar-scaled chemical composition. Continuum flux at the same wavelengths were 
also considered for ATLAS low-resolution spectra (sampled at 20 \AA) of solar abundance. The comparison of these 
fluxes is shown in Fig.~\ref{fig:compar_NextKur_flux1}, for $\log g=4.0$ and for $3000 \le T_{\rm eff} \le 7500$ K, which are a typical gravity and 
temperatures of low-mass members of star-forming regions. At all temperatures, the flux obtained with the GAIA model is lower 
than the flux obtained using ATLAS models, but for $T_{\rm eff} \ltsim 4400$ K (depending on the line wavelength) the 
flux decrement of GAIA spectra is more pronounced than the ATLAS spectra. This is particularly evident, for instance, at 
$\lambda=$ 5600, 6300 \AA, and is due to the formation in stellar spectra of molecular bands, such as metal oxide (most of all TiO, 
but also VO), hydroxide (such as OH), hybrids (such as CaH, FeH, MgH) in the visible, and CO and H$_2$O in the near infrared. 
These bands are not accurately reproduced by ATLAS models, which lack line opacity computations for both triatomic 
molecules (with the exception of H$_2$O) and numerous diatomic molecular transitions (such as VO). 

Since the lines used for abundance measurements are spread over wide spectral ranges (typically in the $4000-8000$ \AA~range), 
we also calculated the synthetic fluxes in the Johnson $BVRI$-bands by integrating the synthetic ATLAS and GAIA spectra, 
weighted by the Johnson transmission curve of the $BVRI$ filters. The results are shown in 
Fig.~\ref{fig:compar_NextKur_flux3} for $\log g=4.0$. The shift between the ATLAS and GAIA models also appears 
in the $BVRI$-fluxes, even if it is less evident than the line-continuum fluxes of 
Fig.~\ref{fig:compar_NextKur_flux1}, because of the integration over the band wavelengths.

\begin{figure*}[h]
\begin{center}				    
\includegraphics[width=8.cm]{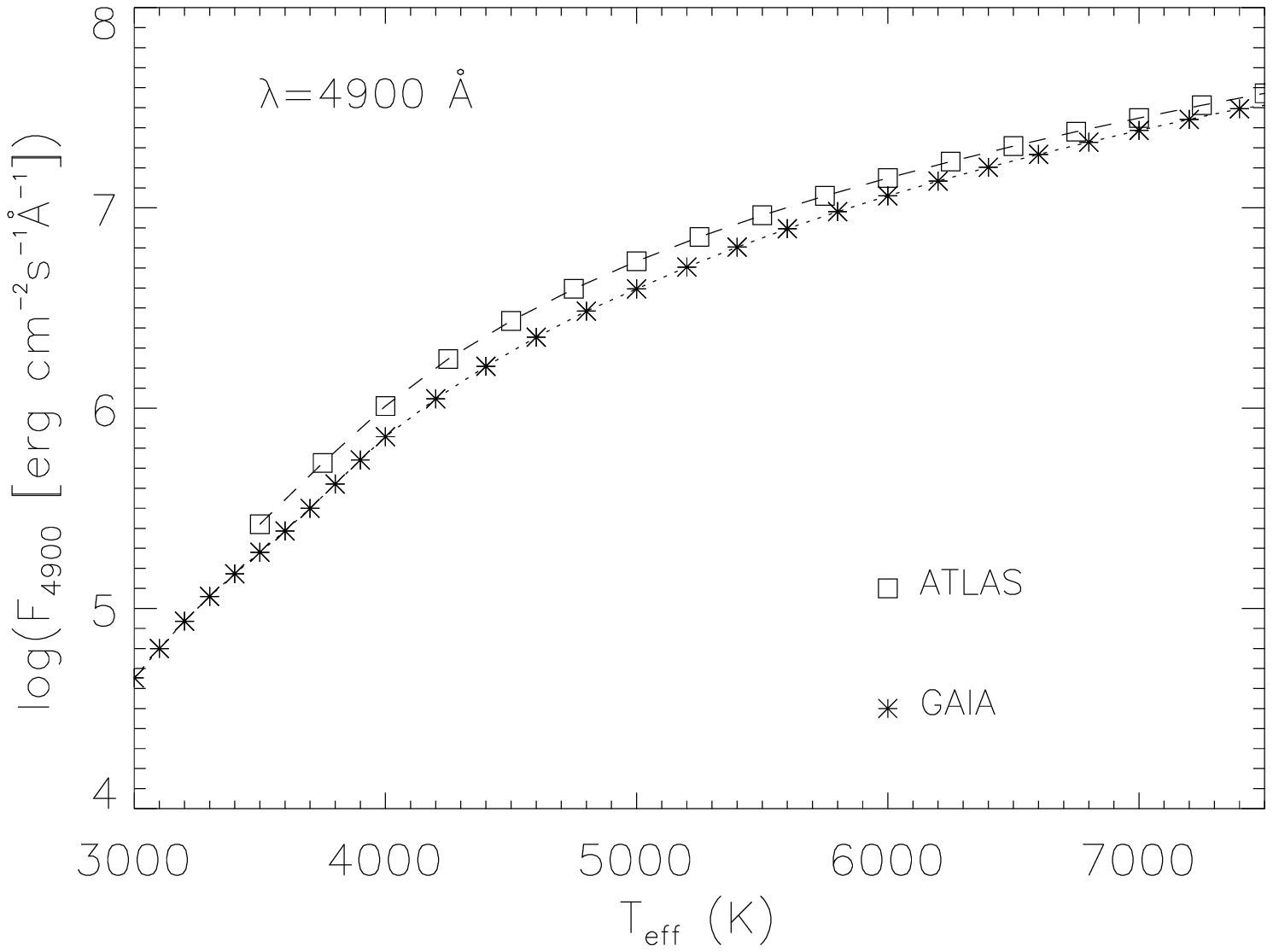}
\includegraphics[width=8.cm]{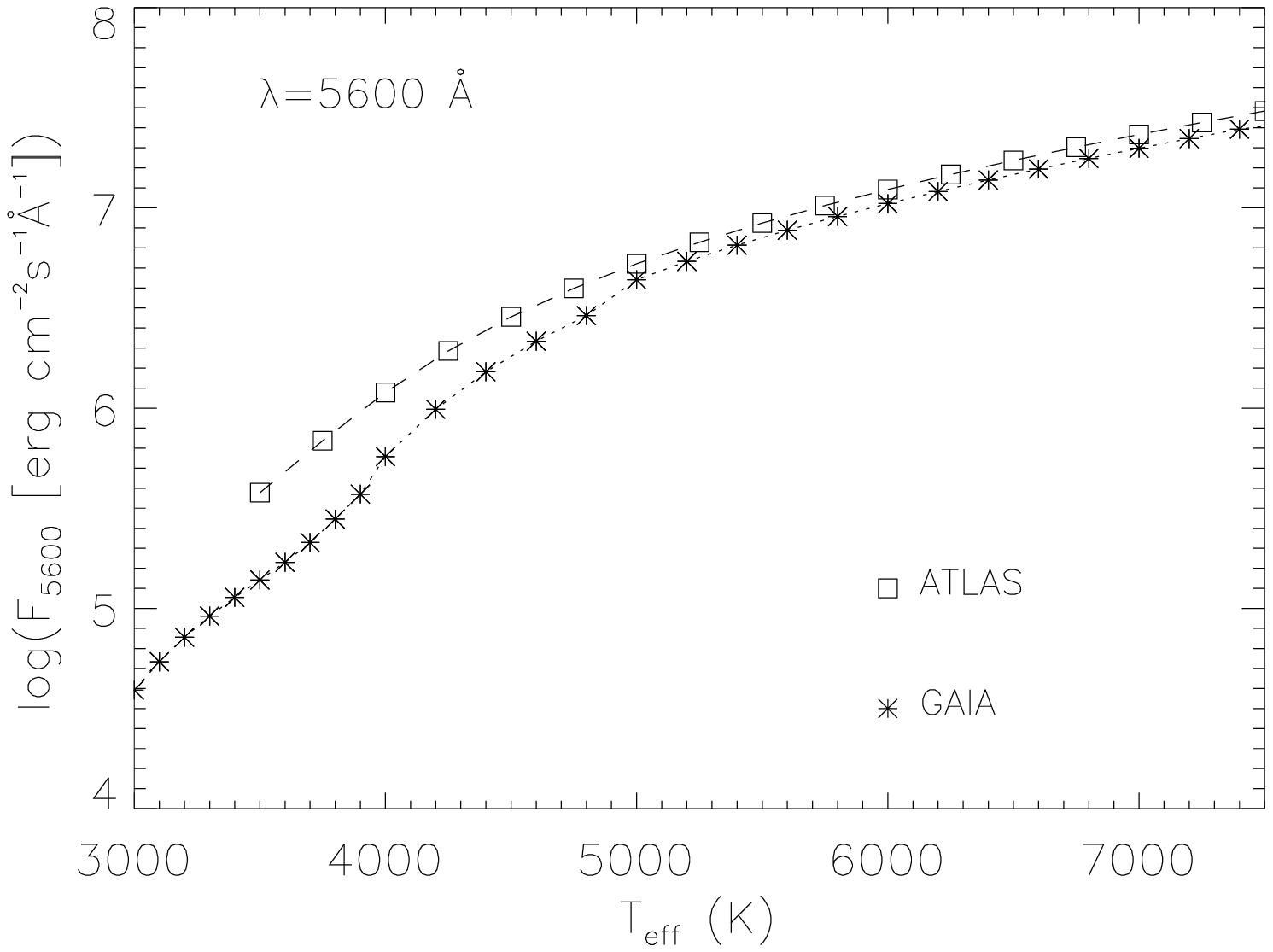}
\includegraphics[width=8.cm]{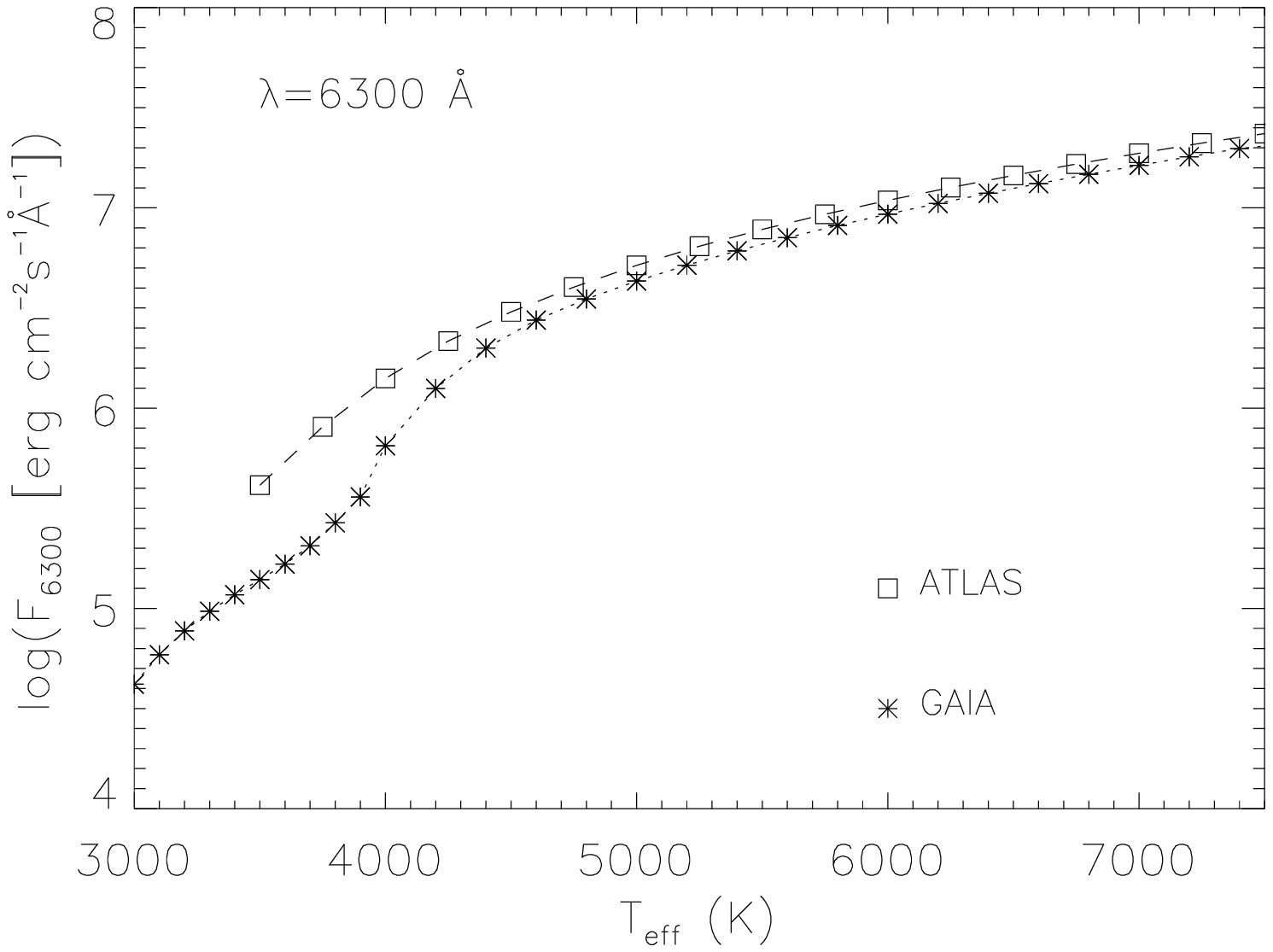}
\includegraphics[width=8.cm]{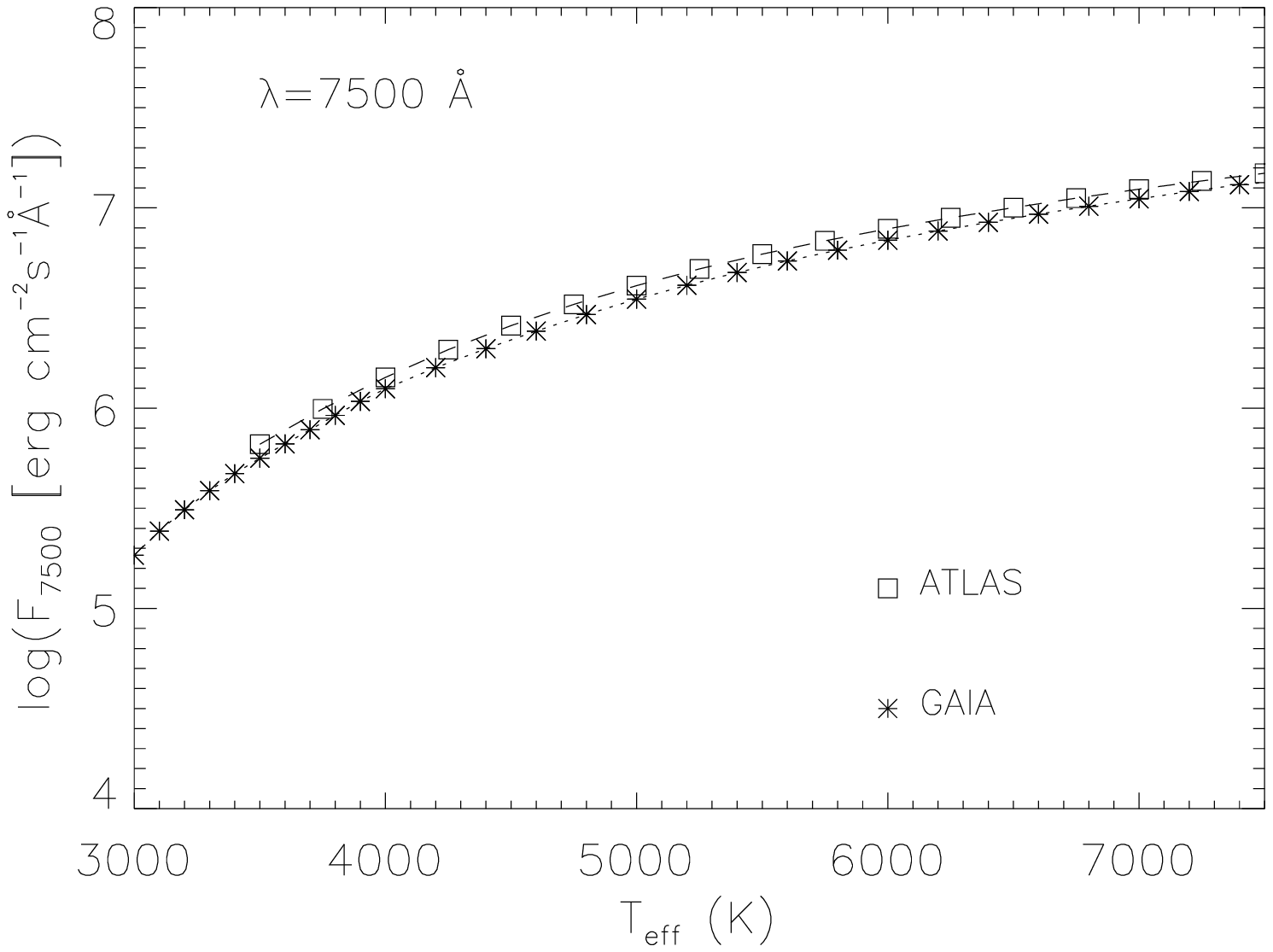}
       \caption{Comparison between continuum flux at $\lambda$4900, 5600, 6300, 7500 \AA~obtained with ATLAS spectra 
       (squares and dashed line) and GAIA spectra (asterisks and dotted line) at $\log g=4.0$ as a function of $T_{\rm eff}$. 
       The lines represent an interpolation through the points.}
       \label{fig:compar_NextKur_flux1}
 \end{center}
\end{figure*}

\begin{figure*}	%[b!]
\begin{center}
\includegraphics[width=8.cm]{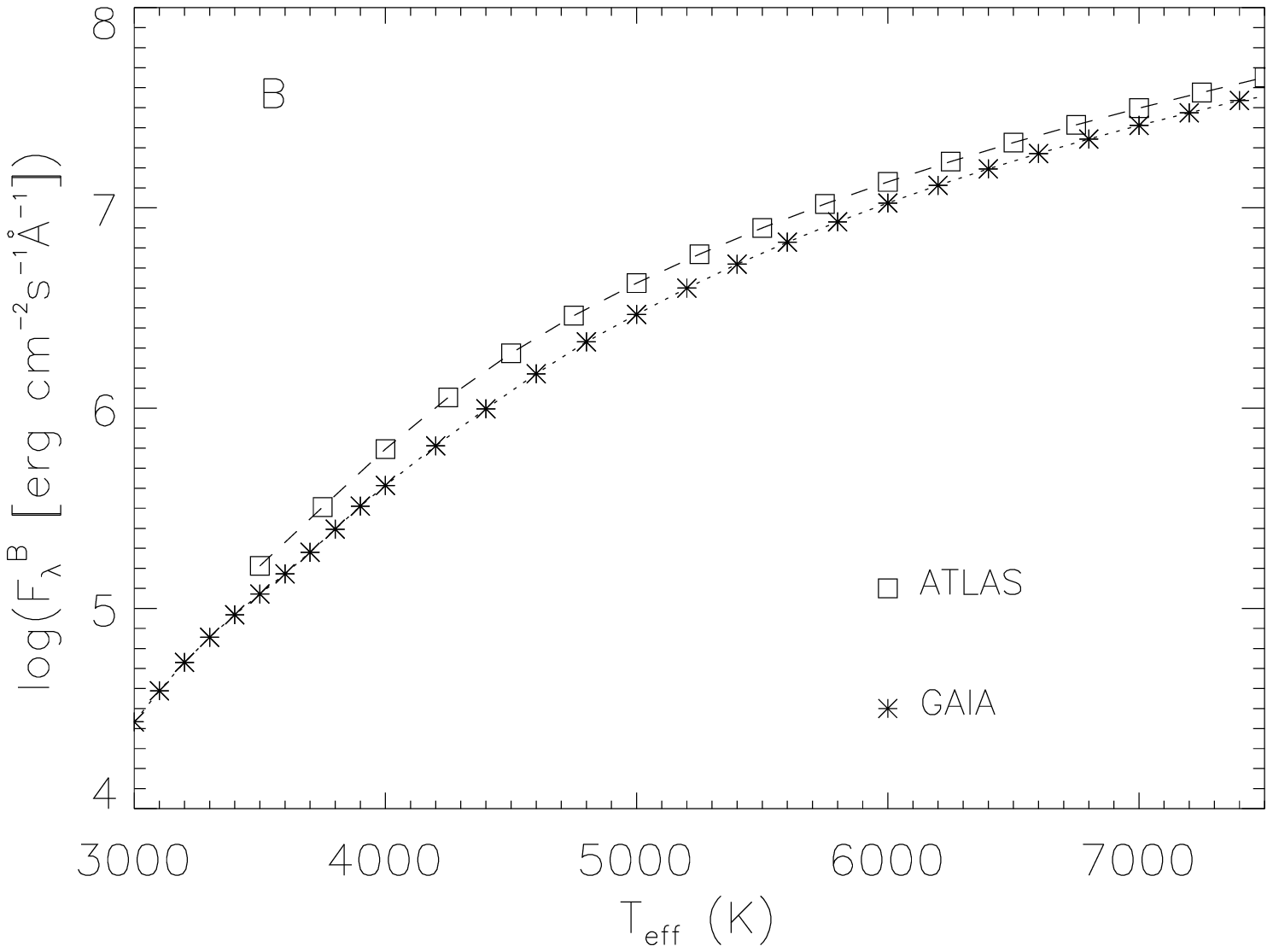}
\includegraphics[width=8.cm]{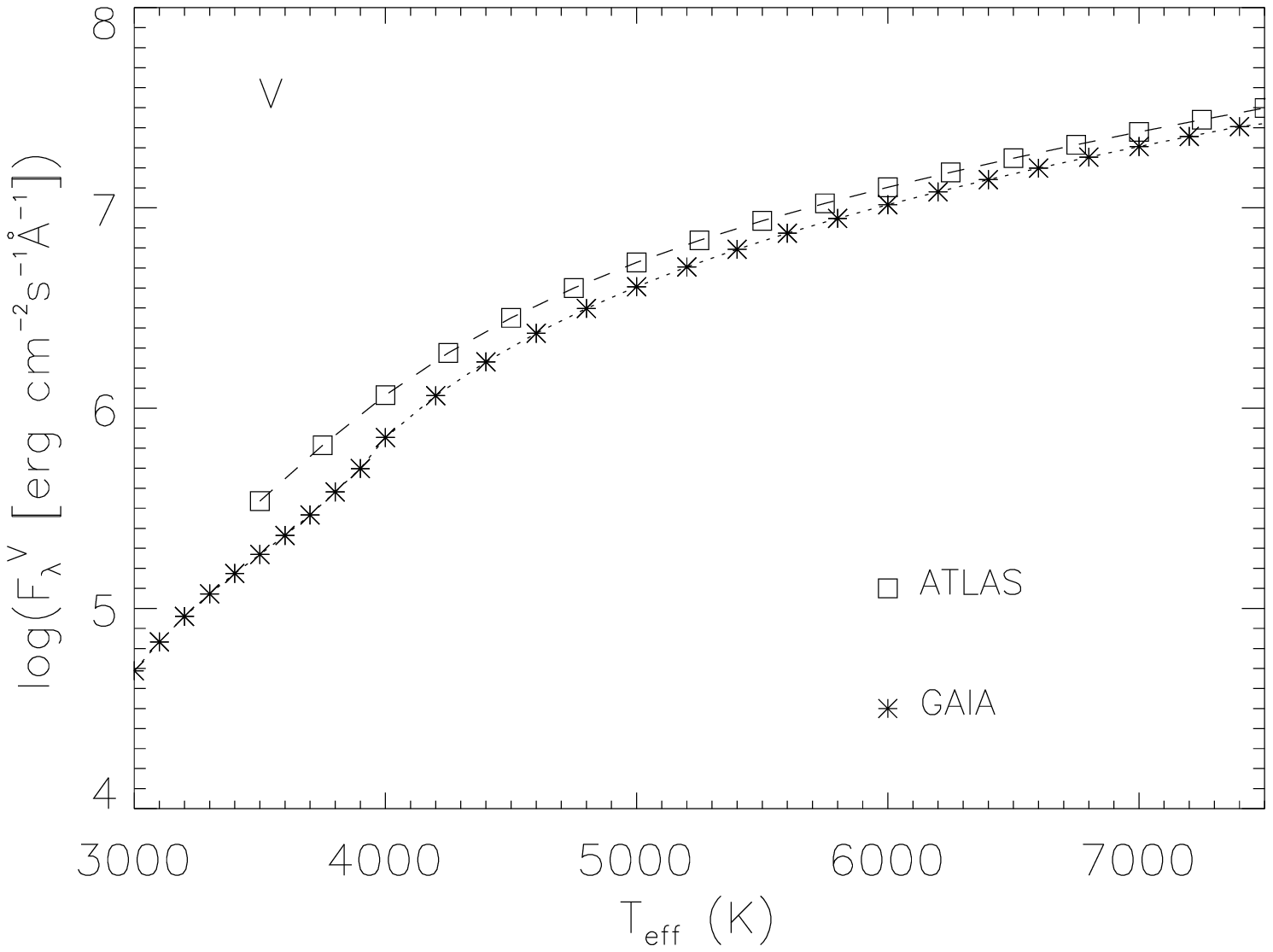}
\includegraphics[width=8.cm]{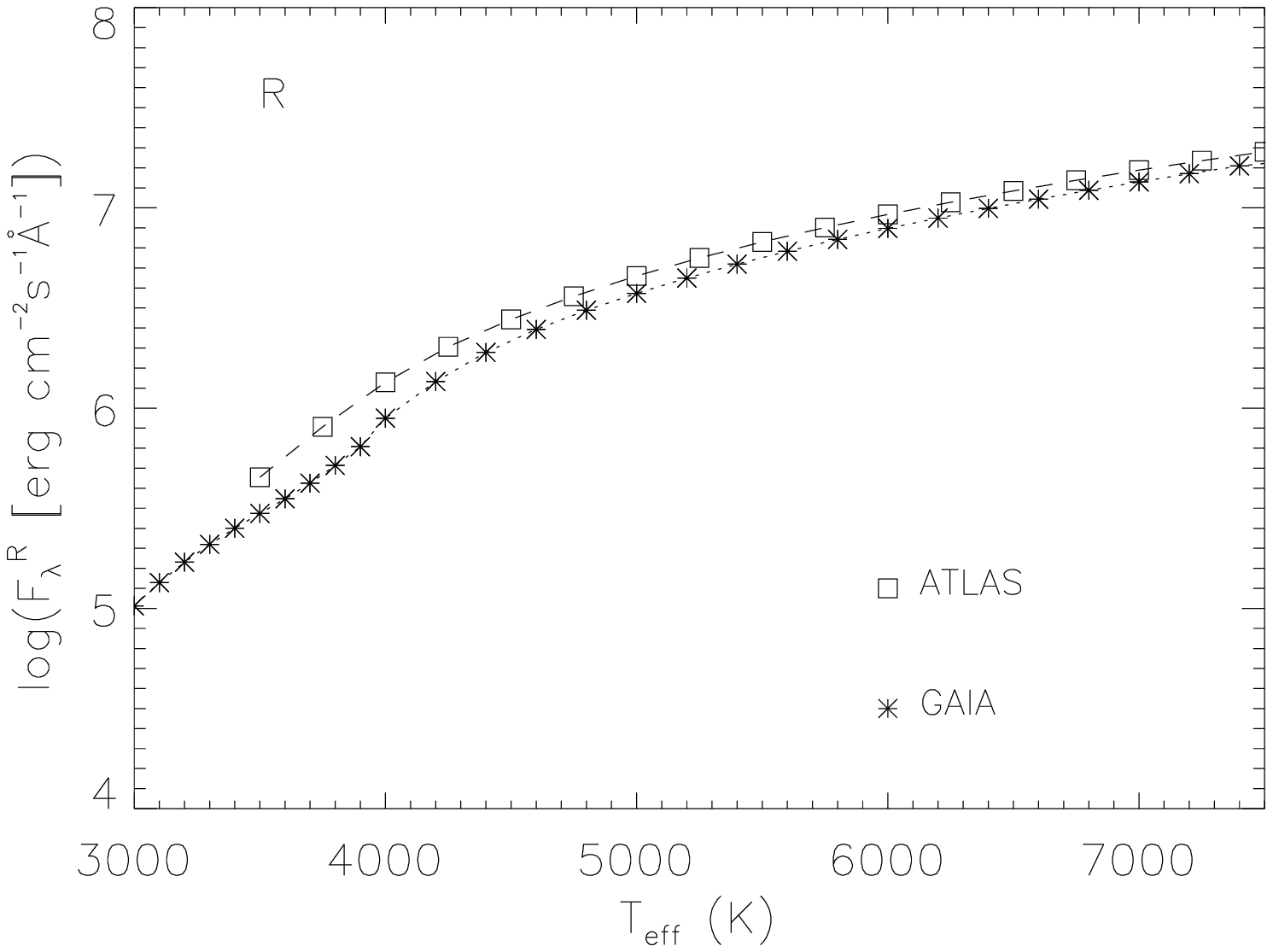}
\includegraphics[width=8.cm]{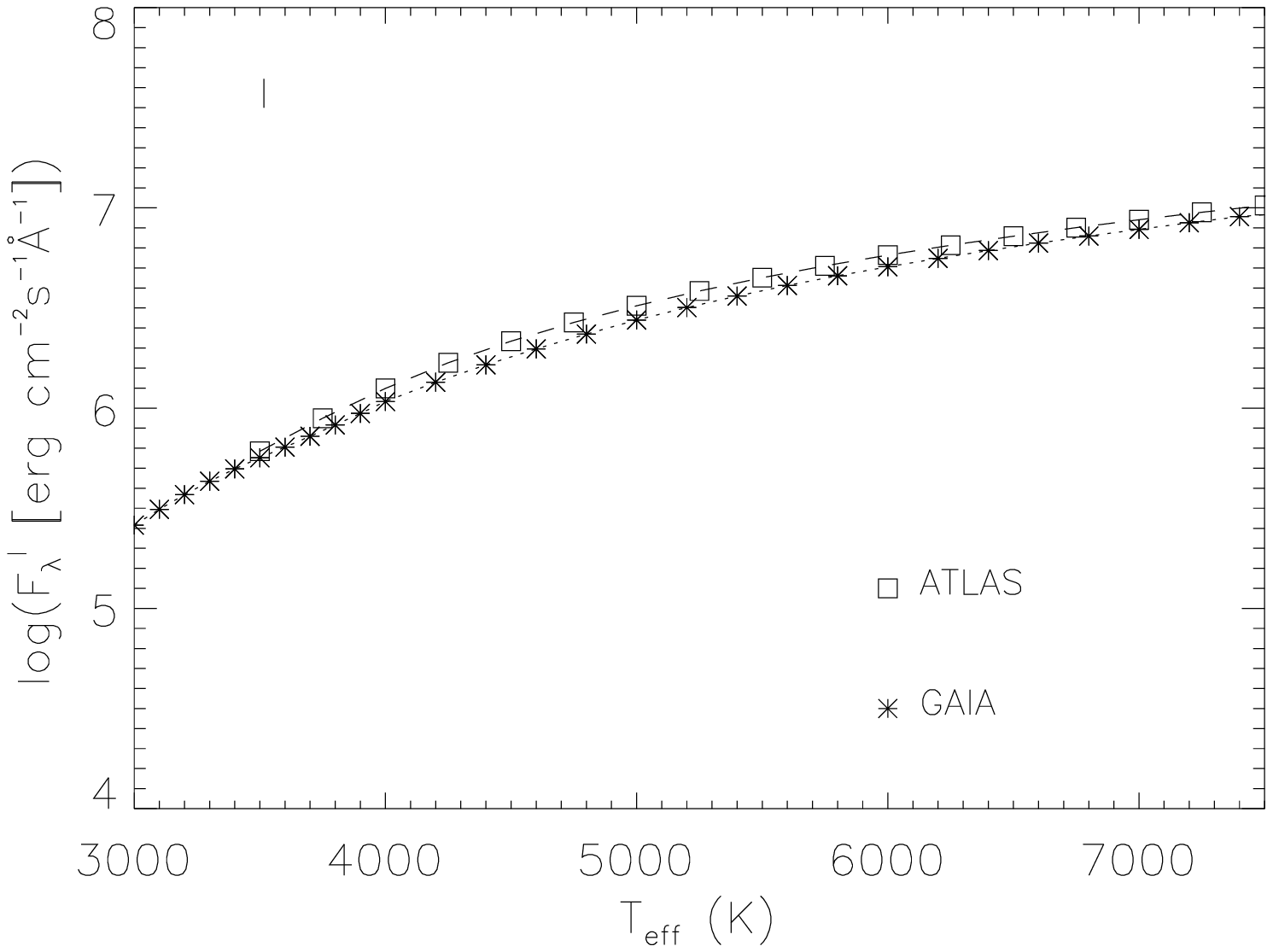}	       
       \caption{Comparison between continuum flux at the Johnson $BVRI$-bands obtained with ATLAS spectra 
       (squares and dashed line) and GAIA spectra (asterisks and dotted line) at $\log g=4.0$ as a function of $T_{\rm eff}$. 
       The lines represent an interpolation through the points.}
       \label{fig:compar_NextKur_flux3}
 \end{center}
\end{figure*}

\subsection{Implications on abundance determination}
\label{sec:abundance_bis}
To search for the effect on abundances, we compared the metallicities and the Na, Al, Si, Ca, Ti, Ni abundances 
obtained using ATLAS and GAIA models for CVSO159, CVSO118, KM~Ori, and the Sun. The results are listed in 
Table~\ref{tab:mean_abun} for both ATLAS and GAIA models. 

\subsubsection{Iron}
The comparison of metallicities is shown in Fig.~\ref{fig:compar_NextKur_abun}, where the difference between 
the two models increases with decreasing temperature, because of the presence of the mentioned bands. In particular, 
at $T_{\rm eff}=4000-4300$ K the difference in abundance is $\pm 0.07-0.08$ dex, while for the Sun the difference 
is only $\pm 0.01$ dex (see also Table~\ref{tab:mean_abun}). As a consequence, for a differential abundance analysis 
with respect to the Sun, for KM~Ori (at 4700 K) we do not find almost any difference between the two models, 
while for both CVSO159 (at 4000 K) and CVSO118 (at 4300 K) the differences are 
[Fe/H]$_{\rm ATLAS}^{\rm CVSO}-$[Fe/H]$_{\rm GAIA}^{\rm CVSO}$=$+0.07$, $+0.08$ dex, respectively. This means that 
the strong departure of the GAIA model from the ATLAS behavior at $\sim4400$ K shown in 
Fig.~\ref{fig:compar_NextKur_flux1} affects iron abundance, leading to similar differences in stars with 
$T_{\rm eff} \ltsim 4400$ K. The lower metallicity resulting from GAIA models (with respect to the ATLAS models) 
depends on the more signidicant formation of molecules in atmospheres with lower temperatures. The molecular 
opacity considered in the GAIA models indeed leads to a redistribution of the flux, which is on average lower 
than the ATLAS one, because of the molecular absorption (line blanketing). Lower flux yields lower intrinsic 
line equivalent widths, which can be reproduced by lower iron abundances.

\subsubsection{Other elements}
In Table~\ref{tab:mean_abun}, we summarize how the use of different models can affect the elemental abundance. 
Here, the comparison between ATLAS and GAIA grids is listed for Na, Al, Si, Ca, Ti, and Ni. 
While lines of elements across over the whole spectrum ($\sim 4800-6800$ \AA) infer very different 
results (such as Ni, besides Fe), elements such as Al, with only two lines at 6696 \AA~and 6698 \AA~close each 
other and not strongly affected by band opacity, lead to similar GAIA and ATLAS abundances.

\begin{table*}  
\caption{Examples of Fe, Na, Al, Si, Ca, Ti, and Ni mean abundances obtained for stars in the Orion complex using 
ATLAS (A) and GAIA (G) models.}
\label{tab:mean_abun}
\begin{center}
\begin{tabular}{c|cccc}
\hline
Name              &  CVSO159 & CVSO118 & KM~Ori  & Sun    \\
\hline
$T_{\rm eff}$ & 4000 K & 4300 K & 4700 K & 5770 K \\
$\log g$ & 3.9  & 4.3  & 3.1  & 4.44  \\
~\\
$<\log n$(Fe)$_{\rm A}>$ & $7.46\pm0.24$ & $7.60\pm0.11$ & $7.37\pm0.07$ & $7.52\pm0.02$ \\
$<\log n$(Fe)$_{\rm G}>$ & $7.39\pm0.23$ & $7.52\pm0.12$ & $7.34\pm0.07$ & $7.51\pm0.02$\\
\\
$<\log n$(Na)$_{\rm A}>$ & $6.04\pm0.07$ & $6.11\pm0.03$ & $6.28\pm0.05$ & $6.31\pm0.04$ \\
$<\log n$(Na)$_{\rm G}>$ & $6.08\pm0.06$ & $6.11\pm0.03$ & $6.26\pm0.06$ & $6.29\pm0.03$ \\
~\\
$<\log n$(Al)$_{\rm A}>$ & $6.27\pm0.04$ & $6.41\pm0.02$ & $6.45\pm0.12$ & $6.48\pm0.03$ \\
$<\log n$(Al)$_{\rm G}>$ & $6.26\pm0.03$ & $6.38\pm0.01$ & $6.42\pm0.14$ & $6.47\pm0.03$ \\
~\\
$<\log n$(Si)$_{\rm A}>$ & ...           & $7.65\pm0.09$ & $7.55\pm0.10$ & $7.56\pm0.03$ \\
$<\log n$(Si)$_{\rm G}>$ & ...           & $7.65\pm0.09$ & $7.51\pm0.11$ & $7.53\pm0.03$ \\
~\\
$<\log n$(Ca)$_{\rm A}>$ & $6.10\pm0.04$ & $6.17\pm0.11$ & $6.22\pm0.10$ & $6.35\pm0.03$ \\
$<\log n$(Ca)$_{\rm G}>$ & $6.12\pm0.03$ & $6.16\pm0.11$ & $6.17\pm0.10$ & $6.34\pm0.03$ \\
~\\
$<\log n$(\ion{Ti}{i})$_{\rm A}>$ & $4.45\pm0.23$ & $4.70\pm0.07$ & $4.77\pm0.08$ & $4.97\pm0.02$ \\
$<\log n$(\ion{Ti}{i})$_{\rm G}>$ & $4.49\pm0.22$ & $4.68\pm0.08$ & $4.74\pm0.09$ & $4.97\pm0.02$ \\
~\\
$<\log n$(Ni)$_{\rm A}>$ & $6.20\pm0.03$ & $6.30\pm0.11$ & $6.01\pm0.10$ & $6.26\pm0.03$ \\
$<\log n$(Ni)$_{\rm G}>$ & $6.11\pm0.03$ & $6.22\pm0.10$ & $6.02\pm0.11$ & $6.24\pm0.02$ \\
\hline		
\end{tabular}
\end{center}
\end{table*}

\begin{figure*}	%[b!]
\begin{center}
\includegraphics[width=9cm]{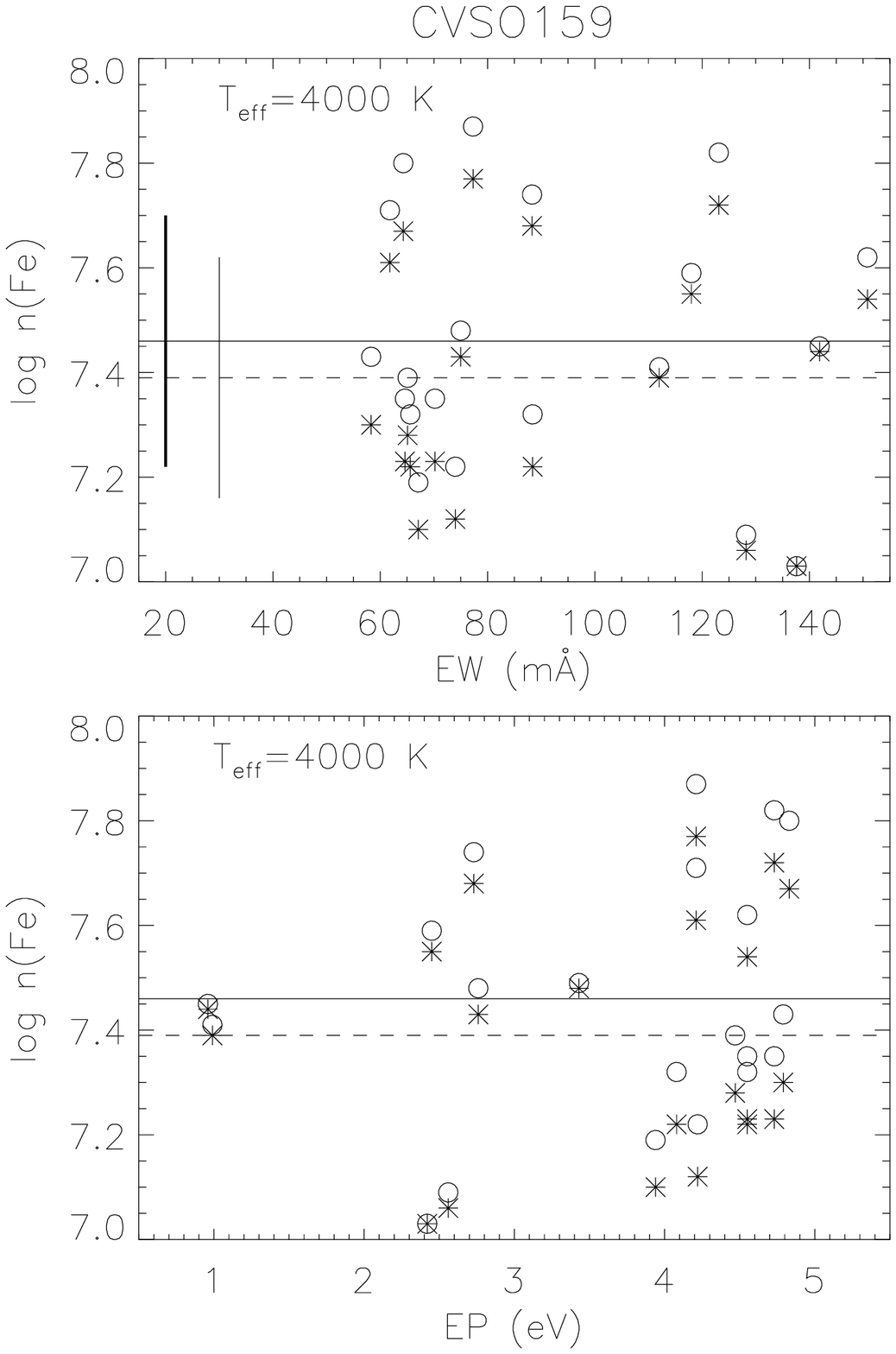}\vspace{.4cm}
\includegraphics[width=9cm]{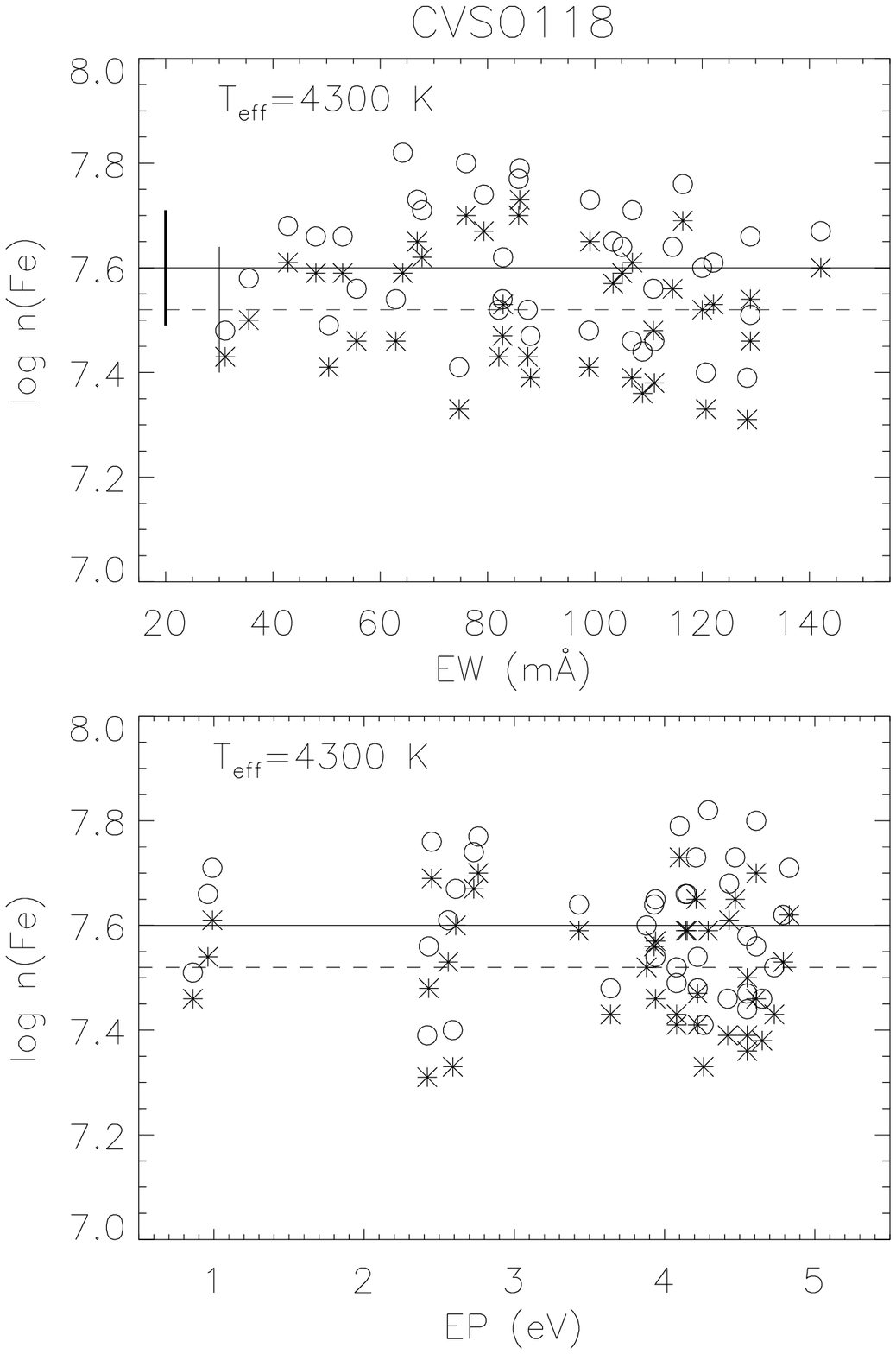}\vspace{.4cm}
\includegraphics[width=9cm]{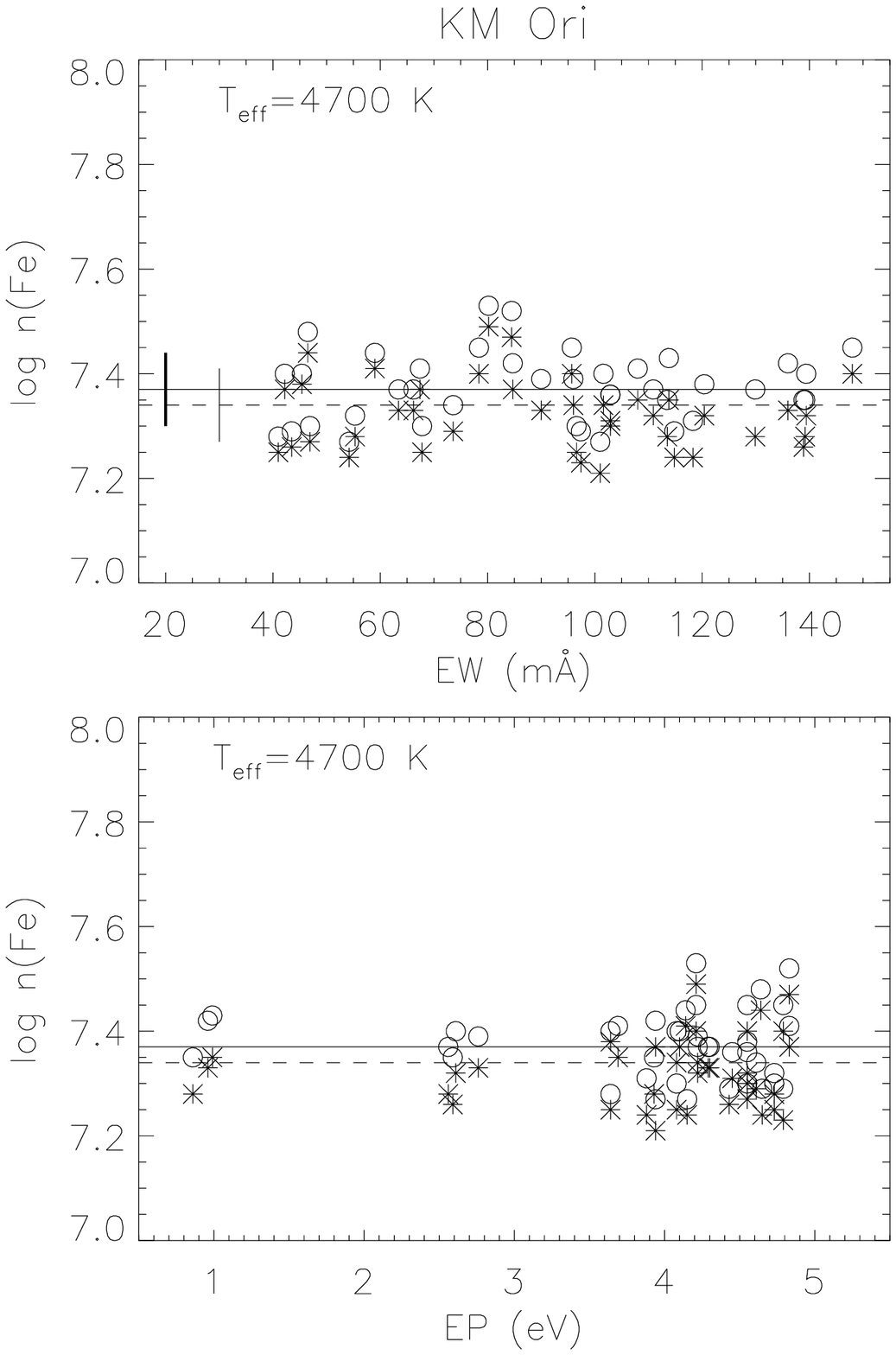}
\includegraphics[width=9cm]{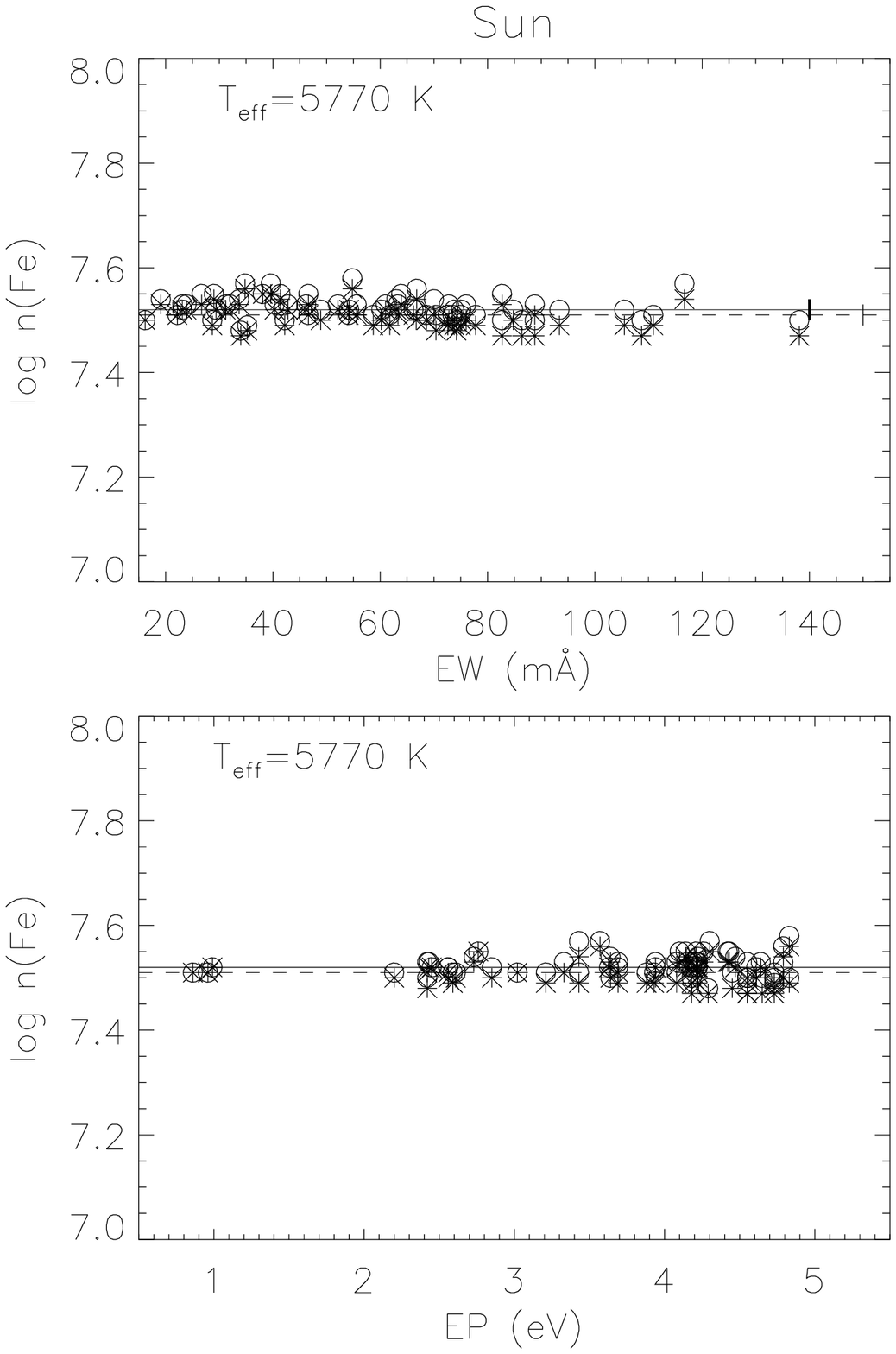}                 
       \caption{Comparison of iron abundances derived by using ATLAS and GAIA model atmospheres as a function of 
       equivalent width (EW) and line excitation potential (EP). These examples display the results obtained for stars with four 
       different temperatures. Circles and asterisks refer to abundances 
       derived with ATLAS and GAIA models, respectively, while solid and dashed lines represent their mean values. 
       Vertical thick and thin bars in the $\log n{\rm (Fe)}$ vs. EW panels are the standard deviations around 
       the average iron abundances.}
       \label{fig:compar_NextKur_abun}
 \end{center}
\end{figure*}

\subsection{Concluding...}
\label{sec:conclusion_bis}
We find $T_{\rm eff} \approx 4400$ K to be the lower limit where the models in which the line opacity computations 
are not fully treated, such as ATLAS, can be applied in an abundance analysis. This has been demonstrated for both 
iron abundance, typically derived by many lines, and other elements ($\alpha$- and iron-peak elements) typically 
used as tracers of chemical enrichment. 

}
\end{document}